\documentclass{aa}  

\usepackage{graphicx}
\usepackage{txfonts}
\makeatletter
\@fleqnfalse
\makeatother                           
\usepackage{orcidlink}                                

\begin{document}

 \title{Solar jet-induced perturbation propagating through coronal loops and in-loop electron beam transport indicated by type II and type N radio bursts}


   \author{
    Yingli Cui\inst{1}\orcidlink{0009-0009-9306-4022}
    \and Xiangliang Kong\inst{1,2}\orcidlink{0000-0003-1034-5857}  
    \and Zhentong Li\inst{3}\orcidlink{0000-0002-4230-2520}
    \and Bing Wang\inst{1}
    \and Yadan Duan\inst{4}\orcidlink{0000-0001-9491-699X}
    \and Ze Zhong\inst{2}\orcidlink{0000-0001-5483-6047}
    \and Hao Ning\inst{2}\orcidlink{0000-0001-8132-5357}
    \and Zhao Wu\inst{1}\orcidlink{0000-0002-6985-9863}
    \and Manqing Wang\inst{1}
    \and Yang Liu\inst{1}\orcidlink{0000-0002-8220-3439}    
    \and Feiyu Yu\inst{1}\orcidlink{0000-0002-1576-4033}
    \and Zelong Jiang\inst{1}\orcidlink{0000-0002-4520-2170}
    \and Wei Chen\inst{3}
    \and Yang Su\inst{3}\orcidlink{0000-0002-4241-9921}
    \and Yao Chen\inst{1,2}\orcidlink{0000-0001-6449-8838}
    }
    
    \institute{
        School of Space Science and Technology, Institute of Space Sciences, Shandong Key Laboratory of Space Environment and Exploration Technology, Shandong University, Weihai, Shandong 264209, People's Republic of China\\
    \email{kongx@sdu.edu.cn}    
    \and 
        Institute of Frontier and Interdisciplinary Science, Shandong University, Qingdao, Shandong 266237, People's Republic of China    
    \and 
        Key Laboratory of Dark Matter and Space Astronomy, Purple Mountain Observatory, CAS, Nanjing 210023, People's Republic of China
    \and 
        Yunnan Observatories, Chinese Academy of Sciences, Kunming, Yunnan 650216, People's Republic of China
    }

 
  \abstract
   {}
   {Solar type II radio bursts are commonly attributed to coronal shocks driven by coronal mass ejections (CMEs). However, some metric type II bursts have occasionally been reported to occur in the absence of a CME and to be associated with weak solar activities. This study aims to identify the driver of the coronal shock in this kind of type II event.} 
   {We investigate a high-frequency metric type II burst with clear band splitting, observed simultaneously by the Chashan Broadband Solar radio spectrograph and the Nançay Radioheliograph. It is associated with a C3.1-class flare and a small-scale jet, but without a detectable CME in the coronagraphs.}
   {The type II burst is preceded by multiple type III bursts, one of which exhibits characteristics of a type N burst. The type II burst source is associated with the jet-induced perturbation front propagating through nearby closed loops at a speed of $\sim$880 km s$^{-1}$, rather than the much slower jet front. This suggests that the disturbance initiated by the jet can convert to a shock wave within low Alfvénic coronal loops, providing the necessary conditions for electron acceleration and subsequent radio emission. Our findings offer new insights into the formation mechanism of high-frequency type II bursts associated with weak flares and jets.}
   {}


  \titlerunning{Solar jet-induced type II radio burst}
   \maketitle

%

\section{Introduction} \label{sec:intro}

Solar flares and coronal mass ejections (CMEs) are intense energy release activities in the solar atmosphere, capable of accelerating a large number of electrons to high energies. These energetic electrons further generate various types of radio bursts through coherent radiation mechanisms, such as plasma emission \citep{1958AZh....35..694G,2008A&ARv..16....1P}. Solar radio bursts at decimeter-meter wavelengths serve as one of the primary diagnostics for the acceleration and transport of energetic electrons in the corona. 

Type II and type III radio bursts are the most commonly observed radio emissions associated with explosive solar activities. Type II bursts appear as narrow bands that slowly drift in radio dynamic spectra and are attributed to energetic electrons accelerated at outward-propagating shock fronts \citep{1985srph.book..333N}. 
Most type II bursts exhibit intermittent or patchy emission bands. In many events, the fundamental and harmonic bands of type II bursts can further split into sub-bands, known as the `band splitting' phenomena \citep[e.g.,][]{1975ApL....16...23S,2001A&A...377..321V,2014ApJ...793L..39D,2015ApJ...812...52D}.
In addition to the main emission lanes, various fine structures are often observed, including herringbones, characterized by short-duration drifts towards low and high frequencies from a type II burst backbone \citep{1983ApJ...267..837H}. 
In contrast, type III bursts drift much more rapidly in radio dynamic spectra, representing the propagation of fast electron beams ($\sim$0.1$-$0.3 $c$, $c$ is the speed of light) accelerated through magnetic reconnection during solar flares and propagating along open or large-scale closed magnetic field lines \citep{2020FrASS...7...56R}. The solar corona is pervaded by closed magnetic loops. When electron beams propagate along a coronal loop, the frequency drift reverses as they pass the loop top and turn toward the Sun, resulting in type III burst variants, including type J and type U bursts \citep[e.g.,][]{2017A&A...606A.141R,2023A&A...669A..28M,2024ApJ...964..108F}. If the electron beams encounter reflection due to the magnetic mirroring effect of coronal loops, a type N burst can be observed \citep[e.g.,][]{1987ApJ...319..503C,2016ApJ...830...37K}.

Metric type II bursts and their fine structures are generally identified as radio signatures of CME-driven shocks through the solar corona \citep[e.g.,][]{2012A&A...547A...6Z,2016ApJ...827L...9F,2017SoPh..292..194L,2019NatAs...3..452M,2021ApJ...913...99K,2024A&A...689A.345K}. However, not all type II radio bursts correlate with CME-driven shocks \citep[e.g.,][]{2011A&A...531A..31N,2012ApJ...746..152M,2015ApJ...804...88S,2023A&A...675A.102K,2024SoPh..299...63V}. 
Different physical mechanisms have been proposed to explain the driver of type II bursts in the absence of associated CMEs. \cite{2015ApJ...804...88S} suggested that when the strongly inclined magnetic loops reconnected with the newly emerging flux near the active region (AR), fast expansion of the loops driven by magnetic tension force could generate coronal shock waves that accelerated energetic electrons to produce type II bursts. \cite{2016ApJ...828...28K} established that the high-frequency type II burst was associated with an extreme-ultraviolet (EUV) wave triggered by magnetic reconnection during the flare, and the wave later evolved into a fast shock as it propagated through nearby loops. 
Specifically, some type II events have been found to be accompanied by only C-class flares and narrow jets \citep[e.g.,][]{2021ApJ...909....2M,2022ApJ...926L..39D,2023ApJ...953..171H,2023A&A...675A..98M}. 
\cite{2021ApJ...909....2M} showed that the type II burst source observed by LOw-Frequency ARray (LOFAR) was located much higher than the jet and propagated much faster. Therefore, they suggested that the narrow jet could produce a piston-driven shock ahead of it. For another event, \cite{2023A&A...675A..98M} found that the type II burst source was ahead of jet-disturbed loops and proposed that the EUV wave may steepen into a shock wave in surrounding regions with low Alfvén speeds.

Recently, a high-performance solar radio spectrograph at meter wavelengths was developed at the Chashan solar radio observatory (CSO). More than 50 type II bursts have been recorded since its operation\footnote{\url{http://47.104.87.104/SRData/CBSm/RadioBurstEvent/typeII/typeIIburst_show.html}}. 
In this paper, we select an event that occurred on 2023 May 8 without accompanying CMEs and was associated with a C-class flare and coronal jet. We investigate the origin of the type II burst by combining multi-wavelength observations.
The paper is organized as follows. We describe the observations in Section~\ref{sec:data} and present the analysis results in Section~\ref{sec:result}. The summary and discussion are given in Section~\ref{sec:summary}.

\section{Observations} \label{sec:data}

The radio burst event that occurred on 2023 May 8 was identified with radio spectral data from the Chashan Broadband Solar radio spectrograph at meter wavelengths (CBSm; \citealt{2024ApJS..272...21C}). CBSm is located at the CSO and is managed by the Institute of Space Science of Shandong University. It is designed to monitor high-frequency solar radio bursts at metric wavelengths. Its frequency range is 90--600 MHz, with a frequency resolution of about 76.294 kHz and a high temporal resolution of up to 0.21 ms. Since its operation in November 2022, more than 50 type II radio bursts have been recorded. 

Simultaneously, the Nançay Radioheliograph (NRH; \citealt{1997LNP...483..192K}) provides radio imaging at ten different frequencies from 150 to 445 MHz for this radio burst event, with a 3-dB bandwidth of 700 kHz. Here we focus on analyzing the NRH data at three frequencies: 150, 173, and 228 MHz. 
The spatial resolution of the NRH is determined by its half-power beam width, calculated using the NRH routine in the SolarSoftWare. During the event, the half-power beam size at 150 MHz was 489.6$\arcsec$ (major axis of beam ellipse) and 264.4$\arcsec$ (minor axis of beam ellipse), at 173 MHz was 420.2$\arcsec$ and 236.6$\arcsec$, and at 228 MHz was 322.1$\arcsec$ and 172.9$\arcsec$, respectively. Due to the early morning observation, the low solar elevation angle resulted in beam elongation. However, since we focus on the variation of the centroids of radio sources, this effect can be negligible. 
Additionally, because NRH radio sources are observed at high frequencies and in the harmonic band, and the event occurred near the solar disk center, positional shifts arising from anisotropic scattering can be neglected for this event \citep[e.g.,][]{2019ApJ...884..122K,2020ApJ...905...43C,2023ApJ...956..112K}.

We examined the coronagraph images in the duration of the type II radio burst, taken by the Large Angle Spectrographic Coronagraph (LASCO; \citealt{1995SoPh..162..357B}) C2 on board the Solar and Heliospheric Observatory (SOHO) and the Sun Earth Connection Coronal and Heliospheric Investigation (SECCHI; \citealt{2008SSRv..136...67H}) on board the Solar Terrestrial Relations Observatory (STEREO), but no signatures of CMEs could be observed. 
The only eruptive feature is the onset of a C3.1-class flare accompanied by a small-scale jet from the NOAA AR 13296 (N13W12). The eruption was observed by multiple instruments, including the Atmospheric Imaging Assembly (AIA; \citealt{2012SoPh..275...17L}) on board the Solar Dynamics Observatory (SDO; \citealt{2012SoPh..275....3P}). The EUV data from AIA are primarily used to analyze the flare and jet in this study. 
The Hard X-ray Imager (HXI; \citealt{2019RAA....19..163S}) payload of the Advanced Space-Based Solar Observatory (ASO-S; \citealt{2019RAA....19..156G}) is employed to observe the non-thermal hard X-ray (HXR) emission of the solar flare. HXI covers an energy range of $\sim$10$-$300 keV for spectroscopy and $\sim$15$-$300 keV for imaging, with a maximum operational temporal cadence of 0.25 s and a spatial resolution of 3.1$\arcsec$ \citep{Su2024HXI}.
Meanwhile, the Gamma-ray Burst Monitor (GBM; \citealt{2009ApJ...702..791M}) on the Fermi satellite provides count rates in the range of 8 keV to 1 MeV.


\begin{figure*}[!htbp]
    \centering
    \includegraphics[width=0.8\linewidth]{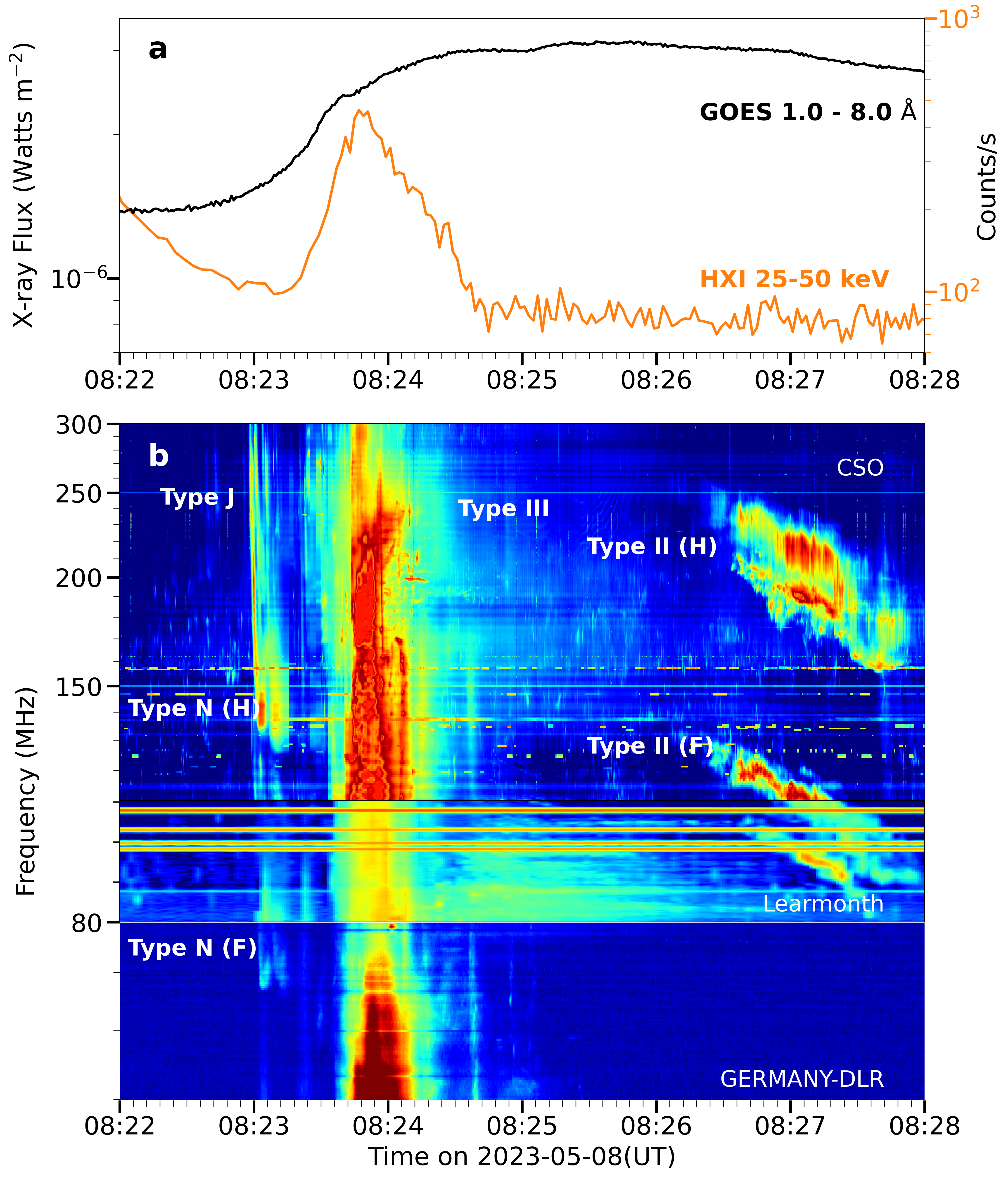}
    \caption{Overview of the event on 2023 May 8. (a) GOES 1$-$8 Å soft X-ray light curve (black) of the C3.1 class flare and the ASO$-$S/HXI count rate  from combined total flux detectors (D92+D93+D94) in the 25$-$50 keV energy channel (orange) between 08:22$-$08:28 UT. (b) Radio dynamic spectrum observed by CSO/CBSm (110$-$300 MHz), Learmonth (80$-$110 MHz), and GERMANY-DLR (50$-$80 MHz), displaying various type III and type III-like (J, N) bursts followed by a type II burst. Both fundamental (F) and harmonic (H) emission lanes can be identified for the type N burst and type II burst. 
    \label{fig1}}
\end{figure*}

\section{Results} \label{sec:result}
A weak C3.1-class flare occurred simultaneously with the type II radio burst, as observed in the GOES soft X-ray (SXR) light curve in Figure ~\ref{fig1}(a). The flare began at 08:20:00 UT, peaked at 08:25:00 UT, and ended at 08:49:24 UT. 
Figure~\ref{fig1}(b) shows a composite of radio dynamic spectrum from CSO/CBSm, Learmonth and GERMANY-DLR between 50$-$300 MHz from 08:22:00 UT to 08:28:00 UT.
Various types of radio bursts, including type III-like (J, N), type III, and type II bursts, were recorded.
During the early impulsive phase of the flare between 08:22:00$-$08:23:24 UT, when the SXR flux began to rise, type III and type III-like radio bursts were already generated, indicating that electrons accelerated by magnetic reconnection had propagated outward along magnetic field lines of different configurations. This results in distinct frequency-drift patterns and intensities on the radio spectrogram.
A type N burst, resembling the letter ``N'' \citep{2016ApJ...830...37K}, occurred before 08:23:00 UT, with other type III bursts around.
An intense period of type III bursts manifested later during the flare impulsive phase between 08:23:30$-$08:24:18 UT, approximately coincident with the peak in the HXR flux.
They drifted continuously to low frequencies (an interplanetary type III burst observed by STEREO-A/WAVES).
The type II burst occurred at 08:26:30 UT after the flare SXR peak time during the gradual phase. It exhibits both fundamental (F) and harmonic (H) lanes, with clear band-spitting in the harmonic emission.

\begin{figure*}[!htbp]
    \centering
    \includegraphics[width=1.0\linewidth]{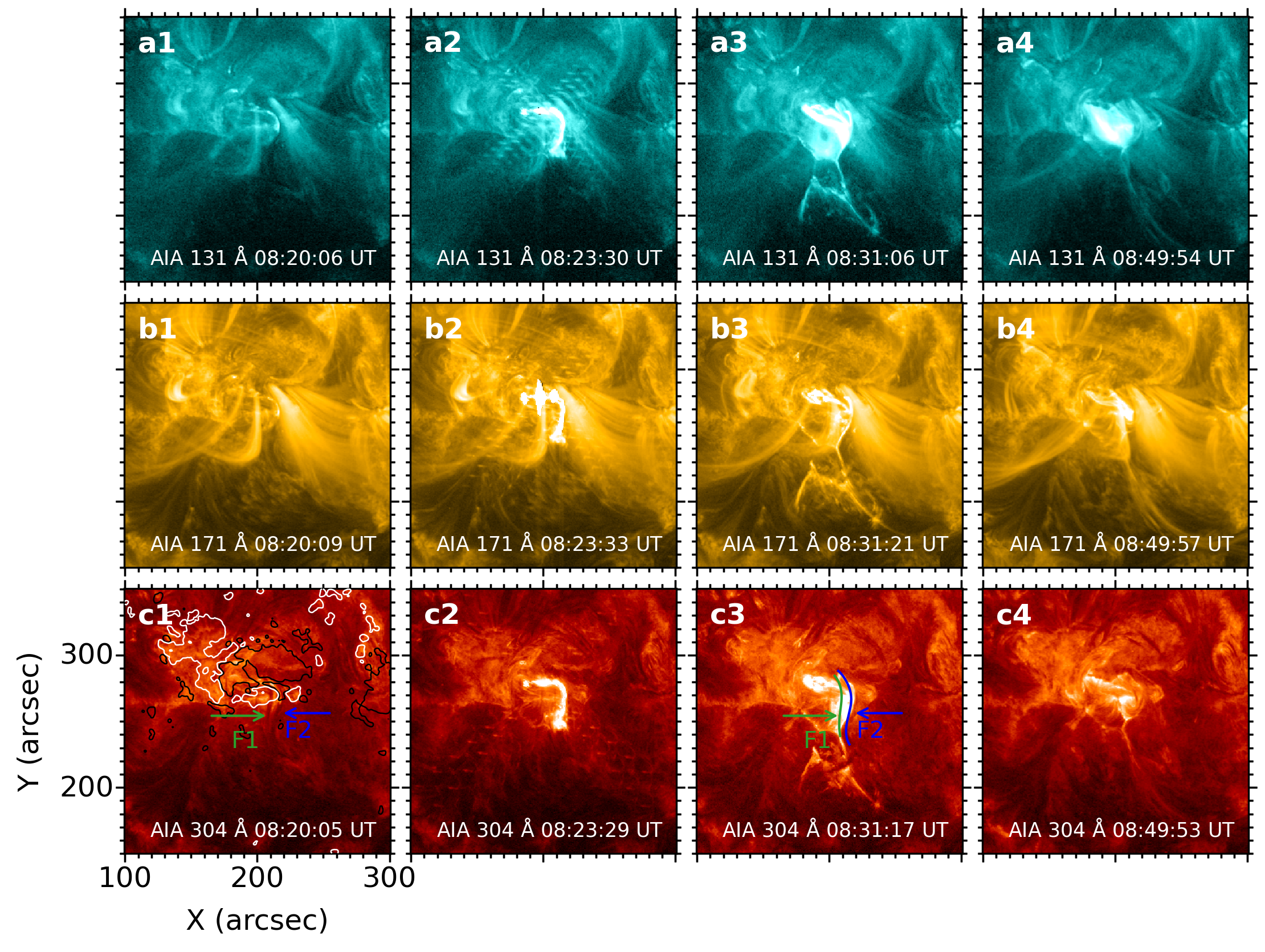}
    \caption{(a1)$-$(a4), (b1)$-$(b4), and (c1)$-$(c4) EUV images in the AIA 131~\AA, 171~\AA\ and 304~\AA\ wavelengths, showing the circular-ribbon flare and jet eruption. White and black contours in panel (c1) represent the positive and negative magnetic fields of the HMI LOS magnetogram scaled at $\pm$120 G, respectively. In panel (c3), green and blue curves mark the locations of filaments F1 and F2, respectively, as observed in panel (c1). 
    \\[0.8em]
    (An animation of this figure is available in the online article.)}
    \label{fig2}
\end{figure*}

\begin{figure*}[!htbp]
    \centering
    \includegraphics[width=1.0\linewidth]{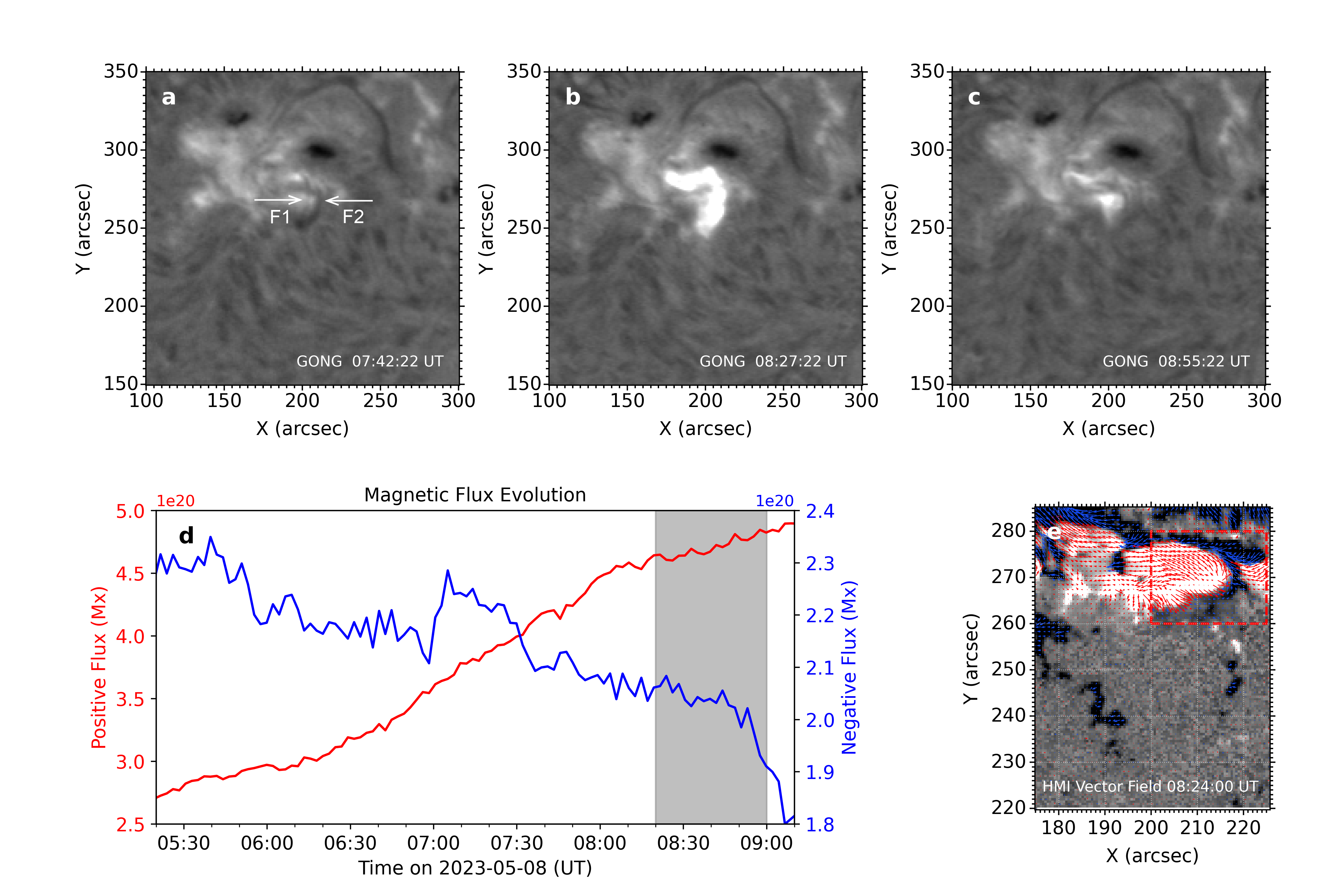}
    \caption{(a)-(c) GONG H$\alpha$ images before, during, and after the jet eruption. White arrows in panel (a) indicate the two mini-filaments. 
    (d) Evolution of the positive (red) and negative (blue) magnetic flux between 05:20-09:10 UT. The gray shaded region indicates the period of the flare. 
    (e) HMI vector magnetic field map at 08:24:00 UT. Red and blue arrows represent the transverse field for positive and negative polarities, respectively. The length of the arrows indicates the magnetic field strength and the direction corresponds to the azimuthal orientation.} 
    \label{fig3}
\end{figure*}

\subsection{Flare and jet eruption} 

The characteristics of the flare and jet in the eruptive AR are shown in Figure~\ref{fig2}. 
Panels (a1$-$a4), (b1$-$b4), and (c1$-$c4) are a sequence of EUV images in the AIA 131 Å,  171 Å and 304 Å channels, respectively, presenting an overview of the eruption process in the AR 13296 near the solar disk center. 
As shown in Figure 2(c1), the AR was located near a magnetic polarity inversion zone, with two clearly visible mini-filaments (F1, F2) in 304 Å. During the early phase of the eruption (08:20:05$-$08:23:09 UT), no clear filament rise can be observed, possibly due to the high altitude of the filaments and intense flare brightening. At 08:23:09 UT, a jet was observed on the southern side of the AR, followed by the ejection of dark filament material with likely untwisting motion (see online animation). After the jet eruption, the mini-filaments disappeared (Figure~\ref{fig2}(c4)).

Figures ~\ref{fig3}(a)-(c) present a series of GONG H$\alpha$ images before, during, and after the jet eruption, respectively. Both filaments F1 and F2 can be identified from the GONG H$\alpha$ data before the eruption. However, filament F1 appears to be indistinct as a result of the relatively lower spatial resolution of GONG. The filament structures finally disappeared after the eruption, as shown in Figure~\ref{fig3}(c), consistent with the observation in AIA 304 Å.

To investigate the trigger mechanism of the filament eruption, we further analyzed the  observation characteristics of magnetic flux and vector magnetic field in the AR. Figure~\ref{fig3}(d) shows the temporal evolution of the magnetic flux near the polarity inversion line (PIL).
It is calculated from the HMI LOS magnetograms for the target region, marked by the red rectangle in Figure~\ref{fig3}(e), over a time range from three hours before the eruption to 10 minutes after the flare (05:20$-$09:10 UT).
The negative magnetic flux exhibits a significant and continuous decrease, while the positive magnetic flux shows a sustained increase due to the movement of positive polarity into the target region, suggesting that magnetic cancellation was ongoing during the flare (08:20$-$09:00 UT, shaded region in Figure~\ref{fig3}(d)).
Therefore, magnetic cancellation may be a key triggering factor in the eruption of mini-filaments. This finding is consistent with the widely observed physical mechanism in quiet-region and coronal hole jets, where small-scale filament eruptions are triggered by magnetic flux cancellation \citep[e.g.,][]{2014ApJ...783...11A,2015Natur.523..437S,2016ApJ...832L...7P,2019ApJ...882...16M}.

Based on the HMI vector magnetograms with a 12-minute cadence, we also analyzed the tranverse magnetic field evolution in the AR. As shown in Figure~\ref{fig3}(e), significant magnetic shear motion is observed near the PIL, where the transverse magnetic field arrows of the positive and negative polarity regions are nearly parallel. Magnetic cancellation and shear cause the opposite polarity magnetic fields to continuously approach each other at the neutral line, triggering magnetic reconnection that alters the magnetic configuration. During this process, the erupting magnetic structure drives both internal and external reconnection: internal reconnection releases significant flare energy, manifesting as a C3.1-class flare with circular-shaped ribbons, while external reconnection, driven by the mini-filaments eruption, opens magnetic field lines to produce the jet \citep[e.g.,][]{2014ApJ...783...11A,2015Natur.523..437S,2016ApJ...832L...7P}.


\begin{figure*}[!htbp]
    \centering
    \includegraphics[width=1.0\linewidth]{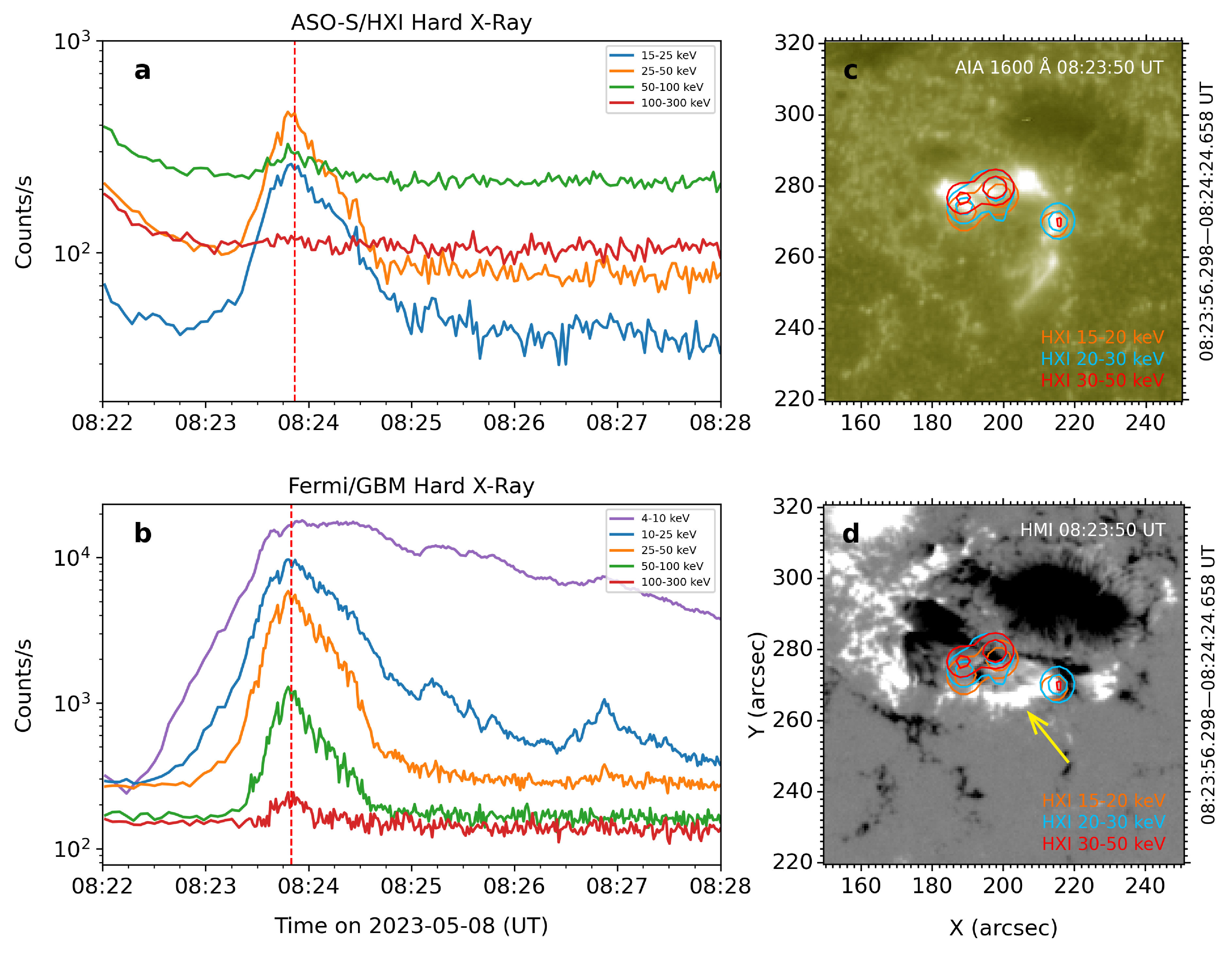}
    \caption{HXR light curves and imaging of the flare in the duration of radio bursts. 
    (a) ASO-S/HXI count rates of combined total flux detectors (D92+D93+D94) in three energy channels between 15$-$300 keV. 
    (b) Fermi/GBM HXR flux in five channels between 4$-$300 keV. The red dashed lines indicate the peak of the curves at 08:23:52 UT. (c$-$d) Contours of HXR sources from HXI, in energy ranges of 15$-$25 keV (orange), 20$-$30 keV (blue), and 30$-$50 keV (red), overplotted on the AIA 1600 \AA\ and HMI magnetogram images. The contour levels represent 25$\%$ and 60$\%$ of the maximum. The yellow arrow indicates the positive polarity in the center of the circular-ribbon.
    \label{fig4}}
\end{figure*}

Energetic electrons accelerated during magnetic reconnection in the jet eruption produced nonthermal HXR emission.
Figure~\ref{fig4}(a) shows that the ASO-S/HXI light curves in the energy range of 15$-$300 keV began to rise around 08:23:00 UT. The flux curves from Fermi/GBM in the energy range of 4$-$300 keV have the same rising trend as the HXI light curves (Figure~\ref{fig4}(b)). Both reach their peaks at 08:23:52 UT, as marked by the vertical red dashed lines. 
The weak GBM signals above 100 keV may originate from the emission of electrons with energies exceeding 100 keV, indicating possible electron acceleration up to $\sim$100 keV in this weak C-class flare. The HXI exhibits no signal above 100 keV, primarily due to its detector area being over 10 times smaller than that of the GBM and its relatively higher background signals.

For this flare event, the HXI is able to provide reliable imaging of HXR sources during the impulsive phase of the flare. 
HXR sources are reconstructed using the background-subtracted counts of HXI detectors from D19 to D91 (which are well calibrated), integrating from 08:23:56 UT to 08:24:24 UT, by HXI\_Clean algorithm \citep{Su2024HXI}. The locations of HXR sources of different energies (in Figure \ref{fig4}(c)) are cospatial with the circular-shaped ribbons in the AIA 1600 \AA\ image, suggesting that the HXR emission is mainly produced at the footpoints of newly reconnected field lines due to the collision of energetic electrons with the low atmosphere. 
As indicated by the arrow in Figure~\ref{fig4}(d), the magnetic polarity in the center of the circular-ribbon is positive and surrounded by negative polarities, consistent with the fan-spine topology\citep[e.g.,][]{2019ApJ...871..105Z,2022ApJ...926L..39D,2024ApJ...968..110D,2022ApJS..260...19Z,2024RvMPP...8....7Z}.
Additionally, as shown in Figure~\ref{fig2}, the shape of the jet as it develops also supports the presence of a fan-spine magnetic configuration in the AR.
Both the flare ribbon and HXR sources are located near the polarity inversion region. 
As shown in Figure~\ref{fig3} and discussed above, magnetic flux emergence and shear motion create preferred sites for magnetic reconnection and electron acceleration in the complex magnetic fields during the jet eruption \citep[e.g.,][]{2012ApJ...754....9G,2018ApJ...866...62C}.

\begin{figure*}[!htbp]
    \centering
    \includegraphics[width=0.9\linewidth]{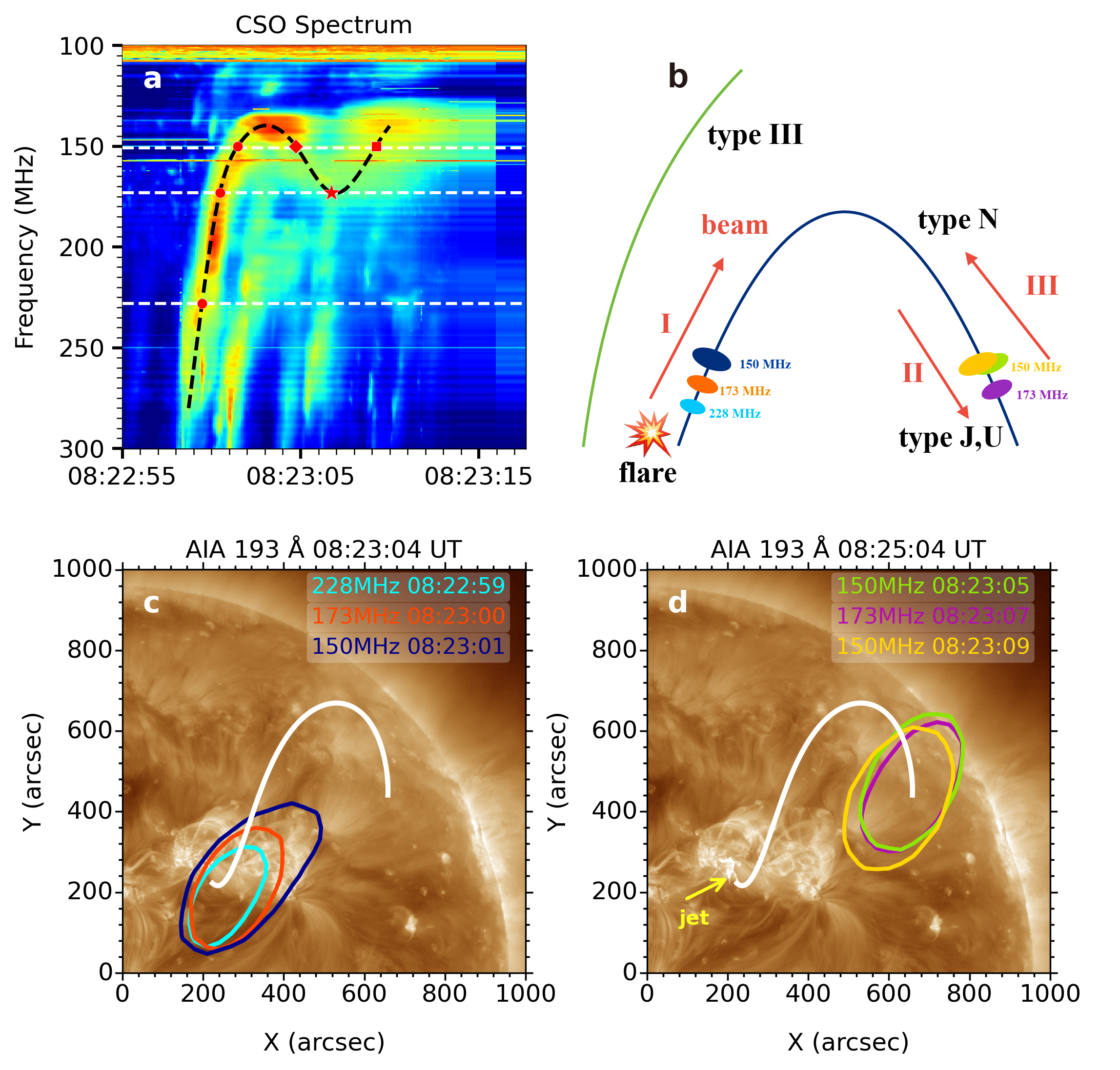}
    \caption{Radio spectral and imaging observations of the type N burst (harmonic emission). 
    (a) CSO/CBSm radio spectrum, the type N burst is outlined by the black dashed curve. White dashed lines indicate the three NRH imaging frequencies at 150, 173, and 228 MHz. Red symbols mark the six selected data points as shown in panels (c) and (d).
    (b) Schematic diagram illustrating the generation of a type N burst. Red arrows indicate the propagation directions of the electron beam, while the blue and green curves represent closed and open magnetic field lines, respectively. The ellipses on the closed loop correspond to the radio source locations shown in panels (c) and (d).
    (c) NRH radio sources superposed on the AIA 193 \AA\ image at three times as marked by red dots in panel (a). The contours represent 70$\%$ of the maximum of each NRH image and are plotted in blue, red, and cyan at three frequencies (150, 173, 228 MHz), respectively. (d) NRH contours corresponding to the data points marked by red diamond, asterisk, and square in panel (a). The white curve represents a schematic of the large-scale closed loop. The yellow arrow points to the position of the jet.
    \label{fig5}}
\end{figure*}

\subsection{Type N radio burst} 

As shown in Figure~\ref{fig1}(b), during the impulsive phase of the flare (before 08:25 UT), multiple episodes of fast-drifting type III and type III-like bursts can be observed, indicating accelerated electron beams propagating outward along open or large-scale closed magnetic field lines. Specifically, the most intense type III burst occurred simultaneously with the HXR peak time. This demonstrates that the electrons generate HXR and radio emissions may have the same origin.
Here we will focus on the likely type N burst in the early impulsive phase. It appeared before 08:23 UT, when the HXR flux in low-energy channels of Fermi/GBM exhibited a rise.
Note that unlike its harmonic counterpart, the fundamental emission of this type N burst is relatively weak and does not exhibit the complete N-shaped profile, consistent with the event investigated in our previous work \citep{2016ApJ...830...37K}. This may be attributed to the fact that the fundamental emission is subject to relatively stronger scattering effects \citep{2019ApJ...884..122K}.

Figure~\ref{fig5}(a) presents a zoomed-in view of the radio dynamic spectrum from CSO/CBSm with a temporal resolution of 0.1 s, highlighting the harmonic emission of the type N burst.
Note that the scale of the frequency-axis is reversed for easier comparison with the schematic diagram illustrating the generation of a type N burst, as shown in Figure~\ref{fig5}(b). The emission lanes of the type N burst are illustrated by the black dashed curve, which consists of three branches. 
The frequency drift rate is approximately -44 MHz s$^{-1}$ from 228 to 150 MHz for the first branch, and 15 MHz s$^{-1}$ from 150 to 173 MHz for the second branch.
\citet{1987ApJ...319..503C} revealed a smaller frequency drift rate for the second branch from the analysis of 16 ``true'' type N bursts, and they explained this phenomenon by the geometrical propagation effect, as the electron beam associated with the second branch is directed toward the Sun.
Other contributing factors include the asymmetry or inclination of coronal loops, energy loss and pitch-angle scattering due to Coulomb collisions, continuous variation in parallel velocity due to the magnetic mirror effect.
Note that the drift rate for the first branch is generally consistent with previous statistical results for type III bursts \citep[e.g.,][]{2018A&A...618A.165Z}, indicating the average speed being $\sim$0.2 c. 
In addition, we can see the duration of emision lanes basically increases for the three successive branches, with the third branch being the most diffuse, which is likely due to the dispersion of the same electron beam along its path. The brightest emission is during the transition between the first and second branches (at the looptop) and the emission generally fades out between the second and third branches (around the mirror point). The above observational properties are generally consistent with those of type N bursts reported in previous studies \citep{1987ApJ...319..503C,2016ApJ...830...37K}.

Unfortunately, we did not identify large-scale closed field lines associated with the type N burst using the PFSS extrapolation model, likely due to the complex magnetic field configuration in the active region. The white curve in Figure~\ref{fig5}(c-d) is drawn based on the direction of the radio source movement and the physical picture of the type N burst. The purpose of using such a shape is to pass through the centroid of the radio sources as much as possible, while accounting for the projection effect.

The first branch is attributed to outward-propagating electron beams, exhibiting negative frequency drift identical to normal type III bursts. It was imaged by NRH at three frequencies, 150, 173, and 228 MHz, as marked by the red dots in Figure~\ref{fig5}(a).  
In Figure~\ref{fig5}(c), the contours (70\% of the maximum) of NRH imaging for the three data points are overlaid on the AIA 193 Å images. It shows that the radio sources for the first branch are located above the AR where the jet erupted. The NRH sources align well with each other at different frequencies. Although the higher-frequency source is expected to be closer to the solar surface, it is hard to determine their altitudes because of the serious projection effect near the disk center.
However, the trend remains apparent that the lower the frequency, the more the radio source tends to shift toward the northwest.
The second branch represents electron beams traversing the apex of large-scale closed loops (as illustrated by the white curve) and undergoing sunward motion. Therefore, the frequency drift reverses and becomes positive. NRH imaged it at the frequency of 150 MHz (the red diamond in panel (a)) and the radio imaging source is shown in Figure~\ref{fig5}(d). It can be seen that the second branch is displaced to the northwest far away from the flare/jet.
The third branch of the type N burst is produced by the electron beams being reflected as a result of magnetic mirror effect of the loops, also imaged at 150 MHz (the red square in panel (a)). The data point around the mirroring point is imaged at 173 MHz at 08:23:07 UT, as indicated by the red asterisk in panel (a).
The locations of the third branch and mirroring point generally align with the second branch, largely owing to the projection effect in the plane of the sky.

From the radio spectrum, we can obtain the crossing time for the electron beam to propagate from the jet side at 173 MHz to the mirroring point located in the northwest, $\sim$7 seconds.
As shown in Figure~\ref{fig5}(c-d), the displacement of the centroids of the NRH sources at 173 MHz is approximately 450$\arcsec$ $\sim$ 315 Mm.
If the propagation path of the electron beam along the closed loop can be simplified as a semicircle, we can estimate the average speed of the electrons to be $\sim$0.24 $c$ (energy $\sim$15 keV).
Due to the magnetic mirror effect in the mangetic loop, the parallel velocity of the electron decreases gradually as it propagates to stronger magnetic field at lower altitudes. This implies that the above estimate can be taken as a lower limit.
The derived beam energy of $\sim$15 keV and beyond is comparable to that of the HXR-emitting electrons, indicating that the type N burst and HXR may be produced by energetic electrons from a similar acceleration process.
The absence of HXR emission at the second footpoint is likely due to the fact that most electrons are reflected, leaving insufficient precipitating electrons to generate detectable emission within the HXI energy range.

\begin{figure*}[!htbp]
    \centering
    \includegraphics[width=0.9\linewidth]{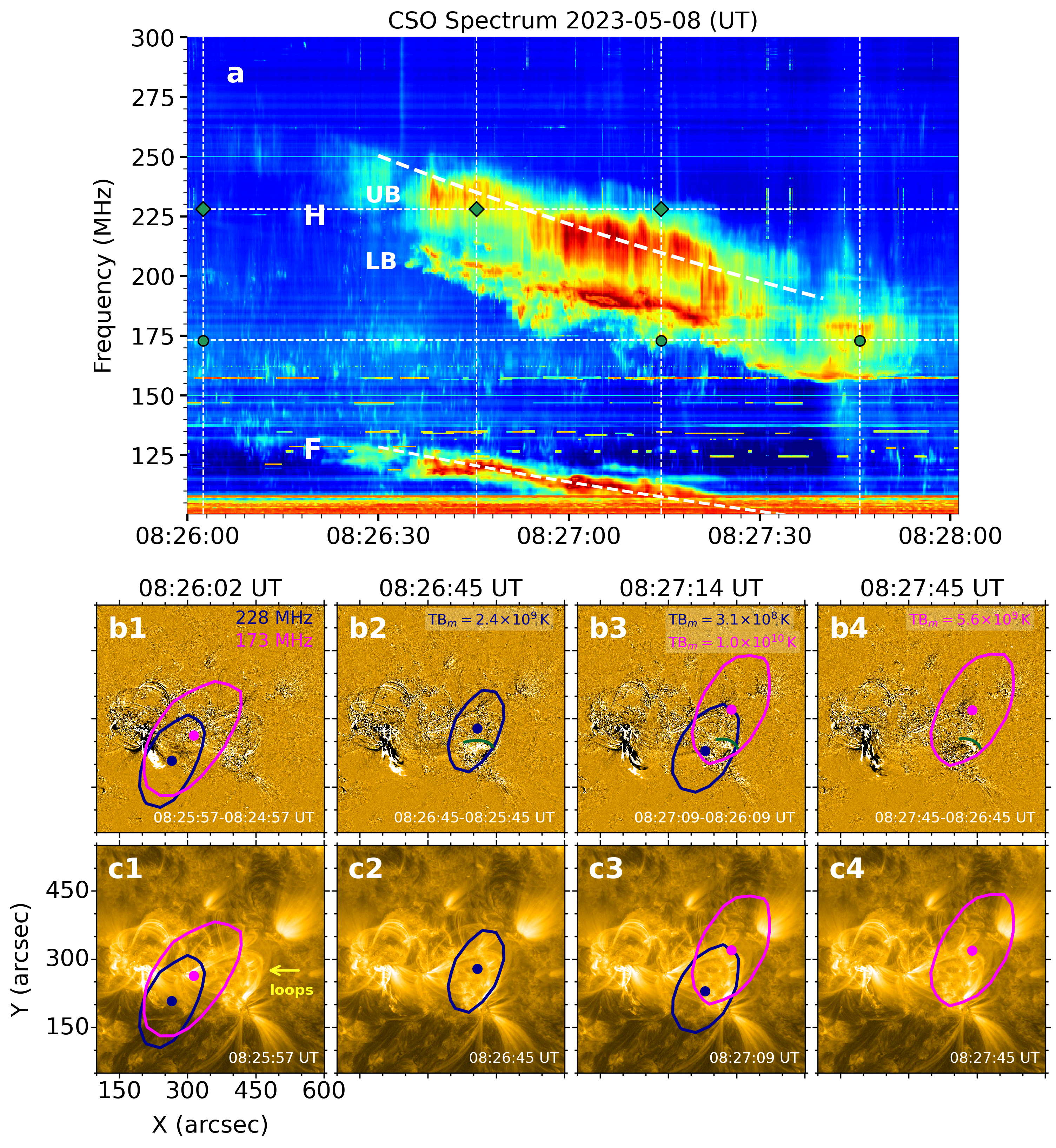}
    \caption{Radio spectrum and temporal evolution of NRH sources of the type II radio burst. 
    (a) CSO/CBSm dynamic spectrum. White horizontal dashed lines indicate the two NRH imaging frequencies, 173 and 228 MHz. Vertical dashed lines indicate the four selected times as shown in the four columns below. Two white dashed curves indicate spectral fitting of the type II burst bands using the Newkirk density model and a shock speed of 800 km s$^{-1}$. 
    (b1)$-$(b4) and (c1)$-$(c4) AIA 171 Å running-difference and original images overplotted with the NRH sources at 173 MHz (magenta) and 228 MHz (dark blue). The contours represent 80$\%$ of the maximum brightness temperatures (TB$_m$) and the filled circles denote the centroids of the radio sources from elliptical Gaussian fits. The green curves in panels (b2)-(b4) highlight the wave-like distrubance front. The yellow arrow in panel (c1) points to the closed loops.
    \label{fig6}} 
\end{figure*}

\subsection{Type II radio burst and its origin} 

The type II radio burst is characterized by both fundamental and harmonic bands with band-splitting and herringbones, as shown in the radio dynamic spectrum in Figure~\ref{fig6}(a). It started around 08:26:20 UT after the SXR peak time and lasted for about 1.5 minutes. The starting frequency is $\sim$125 MHz in the fundamental band and $\sim$250 MHz in the harmonic band. The average frequency drift rate of the harmonic band is about $-$0.53 MHz s$^{-1}$. 
Consequently, that of the fundamental band is $-$0.265 MHz s$^{-1}$. While the drift rate is comparable to that observed in a similar frequency range \citep[e.g.,][]{2015ApJ...804...88S}, it is much faster than the drift rates of type II bursts occurring in lower frequency ranges \citep[e.g.,][]{2015ApJ...812...52D}. The larger frequency drift rates observed at higher frequencies can be attributed to the steeper plasma density gradient in the lower corona, which results in a more rapid decrease in plasma frequency with increasing height.

The harmonic band splits into two sub-bands, the lower frequency band (LB) and the upper frequency band (UB). 
The NRH observed the harmonic band of the type II burst at two frequencies, 173 and 228 MHz, as indicated by the two white horizontal dashed lines Figure~\ref{fig6}(a). 
Note that the NRH did not provide imaging at frequencies between 173 and 228 MHz. Here we use the NRH imaging data with an integration time of 1 s.

We now examine the temporal and spatial evolutions of NRH radio sources relative to the jet eruption. NRH sources were overplotted on the AIA 171 Å running-difference and original images at four selected times, as shown in Figures \ref{fig6} (b1)$-$(b4) and (c1)$-$(c4), respectively.
The maximum brightness temperatures (TB$_m$) of type II sources are indicated in panels (b2)$-$(b4). The centroids of the radio sources are determined by applying an elliptical Gaussian fit to the 70$\%$ contour levels of the TB$_m$. The uncertainty in the centroid positions should not exceed the pixel size, $\sim$30$\arcsec$.
Due to the jet eruption, closed loops west of the AR were disturbed.
However, the disturbance on the western loops is very weak and can not be clearly observed, particularly in the original EUV images.
From the 1-minute cadence EUV running-difference images, we can identify weak signals of the wave-like perturbation front propagating through the closed loops, as highlighted by the green curves in panels (b1)$-$(b4). 
Similar jet-induced perturbation fronts can basically be observed in the running-difference images in 193 Å and 211 Å wavelengths as well.

As shown in Figure~\ref{fig6}(b1), at 08:26:02 UT, before the onset of the type II burst, the radio sources at both frequencies were initially located near the flare and jet. In panel (b2), at the time of the UB backbone, the location of the radio source at 228 MHz shifted to the northwest, coincident with the perturbation front in the EUV running-difference image. Type II bursts are believed to be associated with shock waves. This temporal and spatial coincidence suggests that the observed perturbation induced by the jet eruption is probably a fast-mode shock. 
In panel (b3), the radio source at 228 MHz, representing the herringbones drifting towards high frequencies from the UB, moved towards the southeast. Compared to the UB backbone source, the herringbones are probably located downstream of the shock front, consistent with the physical scenario of herringbones \citep{1983ApJ...267..837H}.
Later, the radio source in the background at 228 MHz moves back to the position of the jet.

The radio sources at 173 MHz are shown in Figures \ref{fig6} (b3) and (b4), representing the LB (likely herringbones) and the UB (at a later time), respectively. Both are cospatial with the disturbed loops. The location of the UB at later times moves further to the northwest, in line with the propagation of the shock over time. The likely LB-associated herringbones, drifting towards lower frequencies from the backbone, are observed to be likely in the upstream of the shock front, as expected. Because the LB emission exhibits patchy bands and is contaminated by the UB emission, NRH lacks the detection of LB backbone emission for this event. The intermittent characteristics and bumps of the type II burst may be attributed to irregular shock fronts as a result of the inhomogeneity of plasma density in closed loops \citep{2015ApJ...798...81K,2023ApJ...952...51K}.

\begin{figure*}[!htbp]
    \centering
    \includegraphics[width=1.0\linewidth]{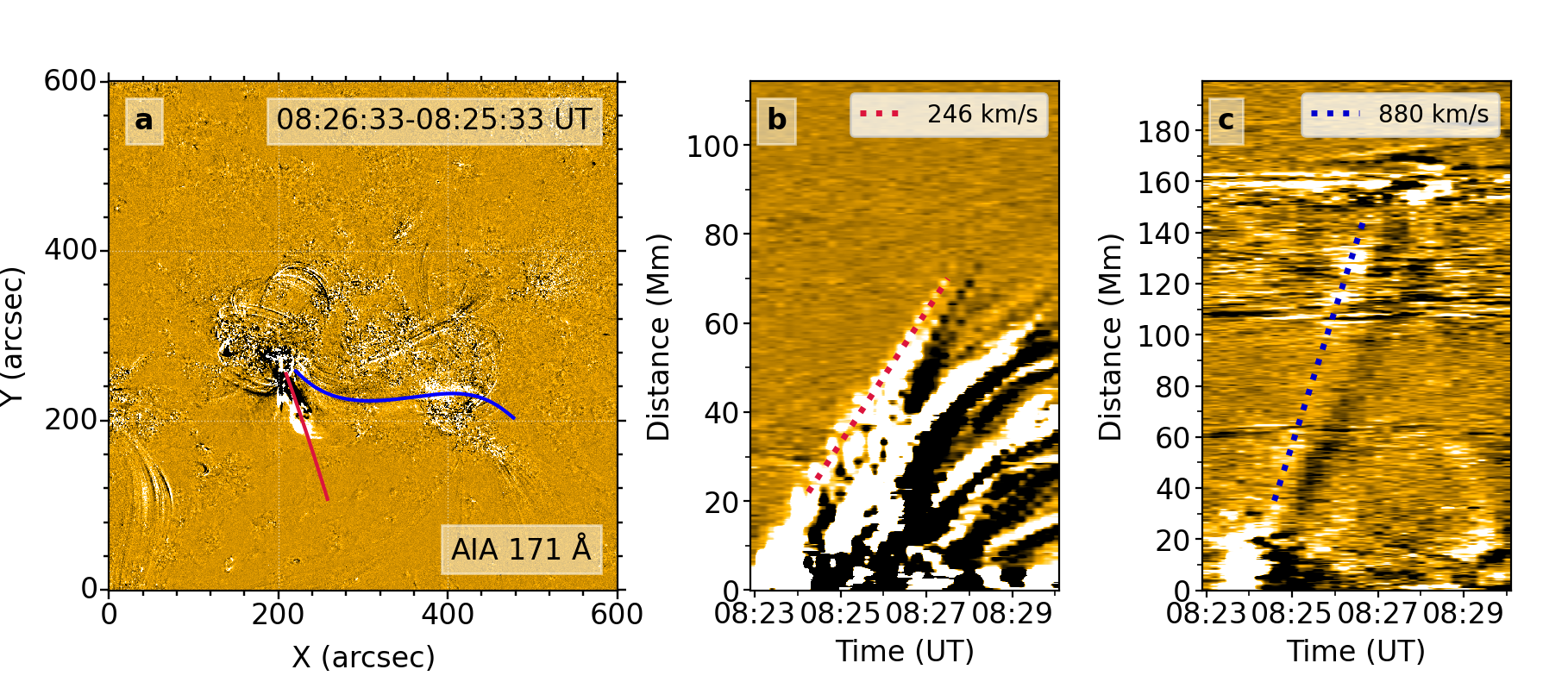}
    \caption{
        Time-distance plots showing the propagation of the jet front and jet-induced disturbance along the loops. 
        (a) Two cuts along the jet front (red) and the closed loop (blue) in the AIA 171 Å running-difference image. 
        (b) The red dashed line indicates a linear fit to the propagation of the jet front. 
        (c) The blue dashed line indicates the propagation speed of jet-induced wave front.      
    }
    \label{fig7}
\end{figure*}

To further explore the origin of the shock wave and type II burst, we analyze the velocities of the jet front and the jet-induced disturbance along the loops. As shown in Figure~\ref{fig7}, we select two cuts in the 171 Å running-difference images to construct the time-distance plots, the red line along the direction of jet eruption and the blue curve along the closed loop. 
The average speeds are derived by applying a linear fit, as indicated by the dashed lines in panels (b) and (c).  The speed of the jet front is $\sim$246 km s$^{-1}$ in the plane of the sky. If we assume the inclination angle of the jet relative to the radial direction is 30$^{\circ}$, the corrected speed will be $\sim$500 km s$^{-1}$. In contrast, the jet-induced perturbation front propagated with a speed of $\sim$880 km s$^{-1}$, significantly faster than the jet front.

We also derived the shock wave speed by performing spectral fitting of the type II burst. Here we apply the Newkik coronal density model \cite{1961ApJ...133..983N},
\begin{equation}
N_e(r)\,(\mathrm{cm}^{-3}) = 4.2 \times 10^{4} \times 10^{\frac{4.32}{r}},
\end{equation}
where $r$ is the radial distance from the solar center in units of $R_\odot$. 
Based on the equation of plasma frequency,
\begin{equation}
f \, (\mathrm{MHz}) = 8.98 \times 10^{-3} \, \sqrt{N_e(r)},
\end{equation}
we found that the type II radio spectrum can be well fitted with the shock speed of $\sim$800 km s$^{-1}$ (see the white dashed curves in Figure~\ref{fig6}), generally consistent with that measured from the time-distance plot of the EUV perturbation front (Figure~\ref{fig7})(c)). 
This provides further evidence that the jet-induced perturbation propagating through nearby closed loops acted as the shock driver of the type II radio burst.
However, it should be noted that this widely used approach relies on the coronal density model, which has significant uncertainties, particularly in the lower corona.

Coronal structures, such as streamers and loops, are characterized by closed magnetic fields and high plasma density, thus have a lower Alfvén speed than the ambient corona. The interaction regions between coronal shocks and these structures have been identified as the source regions of both type II radio bursts \citep[e.g.,][]{2003ApJ...590..533R,2015ApJ...798...81K,2021ApJ...913...99K,2023A&A...675A..98M} and high-energy protons \citep[e.g.,][]{2016ApJ...833...45R,2020ApJ...893...76K,2022ApJ...926..227F,2023ApJ...954..203L}, due to their higher Alfvénic Mach number and the quasi-perpendicular shock geometry.
Using the MHD model, \cite{2023A&A...675A..98M} showed that the type II centroids were located in the region with enhanced density inside the streamer, where the minimum Alfvén speed is only $\sim$100 km s$^{-1}$.
\cite{2005A&A...435.1123W} presented an analytical model of the Alfvén speed in the corona and found regions of minimum Alfvén speed (thus high magnetosonic Mach number) at about 1.2$-$1.4 $R_\odot$, where the Alfvén speed generally $<$500 km s$^{-1}$.
In our event, the starting frqeuency of the harmonic emission of the type II burst is about 250 MHz, therefore we can get the number density $N_e$ $\sim$ 2 $\times$ 10$^8$ cm$^{-3}$ and a height of $\sim$1.2 $R_\odot$ from Equations (1) and (2). Considering the typical magnetic fields in active region loops, $B \sim$2$-$5 G, we can estimate the Alfvén speed is about 310$-$770 km s$^{-1}$. Comparing with the jet-induced perturbation speed $\sim$880 km s$^{-1}$, it yields a fast-mode Mach number 1.1$-$2.8. This suggests that the jet-induced perturbation can steepen into a fast-mode shock wave within
low Alfvénic coronal loops.

\section{Summary and Discussion} \label{sec:summary}

In this paper, we investigate the origin of a type II radio burst recorded by the CSO/CBSm using multi-wavelength observations. The type II burst was associated with a C-class flare and a small jet, in the absence of a CME. During the impulsive phase of the flare, electrons were accelerated through magnetic reconnection and generated HXR footpoint sources cospatial with the circular ribbons and multiple episodes of type III bursts. A type N burst was observed in the early impulsive phase, while the intense type III burst followed by an interplanetary type III burst occurred around the HXR peak time. 
NRH imaging shows that the type II radio sources are cospatial with the wave-like perturbation fronts propagating through nearby loops with a speed of $\sim$880 km s$^{-1}$, instead of being in front of the jet, as observed in EUV running-difference images. This supports the scenario that a shock wave can form in jet-disturbed surrounding regions with a low Alfvén speed, subsequently accelerating electrons to produce the type II burst. Irregular shock fronts due to the density inhomogeneity of the loops can explain the bumps and patchy structures in the type II emission bands.

Currently, there is ongoing debate regarding the origin of coronal shock waves that generate metric type II bursts \citep{2008SoPh..253..215V}. 
While they are often associated with CME-driven shocks, some events may instead be linked to flare-driven blast waves, jet eruptions, or localized reconnection processes.
For the type II event in this study, the solar eruption was a small-scale jet eruption and no clear CME signatures were detected by LASCO/C2 and STEREO/COR1 coronagraphs (angular separation of 9$\degr$). We note that this may be attributed to the active region being located close to the solar disk center, as it is affected by projection effects and the field of view of the coronagraphs.
Previous studies have also reported that some type II bursts lack CME association \citep{2015ApJ...804...88S,2016ApJ...828...28K,2023ApJ...953..171H,2023A&A...675A.102K,2023A&A...675A..98M}. 
This suggests that there may exist a specific category of metric type II events at higher starting frequency, in which the shock drivers could be different from wide and fast CMEs, therefore changing our traditional paradigm of type II bursts. 
Future studies combining newly developed solar radioheliographs in the meter and decimeter wavebands are needed to understand the origin and evolution of coronal shocks.

Despite being associated with a small flare, the event still produced electrons with energies up to $\sim$100 keV, as shown by the Fermi/GBM HXR flux (Figure ~\ref{fig4}).
This significant phenomenon has also been reported in previous studies. 
Recently, \cite{2025ApJ...987L...4C} statistically analyzed 1331 C-class flares recorded by ASO-S/HXI from December 2022 to October 2024. They identified 127 events as high-energy C-class flares, characterized by $>$30 keV HXR emission. The event in our study is precisely one of them, in which the HXR flux of ASO-S/HXI exhibits a signal response up to 70 keV. This indicates that despite relatively weak SXR emission, some C-class flares still possess significant capability in non-thermal electron acceleration. Microflares were shown to efficiently accelerate electrons to high energies under specific magnetic field conditions \citep{2013ApJ...765..143I,2024A&A...691A.172B}. For instance, \cite{2024A&A...691A.172B} noted that these small flares are typically rooted in sunspots, where strong magnetic fields can provide ample energy for magnetic reconnection. However, the specific acceleration mechanism for each individual event requires in-depth analysis.

Our observational results support the fan$-$spine magnetic topology in the active region. Magnetic reconnection can therefore occur both at the coronal null point and near the polarity inversion line where magnetic flux cancellation and shear are observed. During the flare impulsive phase, a fraction of the accelerated electrons propagated outward along the closed filed lines, generating the type N burst and intense type III burst; the other fraction precipitated into the chromosphere, producing the footpoint HXR sources, accompanied by brightening of circular-ribbons in EUV 1600 Å. 
However, with limited observations of this flare event, it is challenging to determine the exact timing and location of magnetic reconnection and electron acceleration.

{Following the onset of the type II radio burst, the Fermi/GBM detected an increase in the 10$-$25 keV energy range around 08:27 UT (Figure~\ref{fig4}), but no corresponding increase was observed by ASO-S/HXI and no associated HXR source was identified. Recently, \cite{2021GeoRL..4895138L} explored the possibility that high-energy electrons accelerated by flare reconnection may be further accelerated by CME-shocks. Through interchange reconnection, some flare-accelerated electrons could move to the vicinity of the shock region and be further accelerated, thereby contributing to the generation of type II radio bursts and solar energetic electron events. Based on the current observations, it remains unconfirmed whether the GBM flux enhancement is related to electrons accelerated by the jet$-$induced shock. As seen from the animation, after the onset of the type II burst, the flare ribbons continued to brighten and expand, indicating that magnetic reconnection was ongoing. Therefore, the increase in GBM flux may be related to the continuous acceleration of electrons by reconnection during the jet eruption process.

\begin{acknowledgements}
    X.K. is supported by the National Key R\&D Program of China under grant 2022YFF0503002 (2022YFF0503000), the National Natural Science Foundation of China under grants 42074203, and the Qilu Young Scholars Program of Shandong University. Z.L. is supported by the Prominent Postdoctoral Project of Jiangsu Province (2023ZB304). D.Y.D. is supported by NSFC grant 12403065. Z.Z. is supported by NSFC grant 12303061, the Shandong Natural Science Foundation of China (ZR2023QA074), and the China Postdoctoral Science Foundation (2023T160385 and 2022M711931). H.N. is supported by NSFC grant 12203031 and the China Postdoctoral Science Foundation (2022TQ0189). Y.S. is supported by NSFC 12333010.
\end{acknowledgements}


\bibliographystyle{aa} 
\bibliography{export-bibtex} 

@ARTICLE{2018A&A...618A.165Z,
       author = {{Zhang}, P.~J. and {Wang}, C.~B. and {Ye}, L.},
        title = "{A type III radio burst automatic analysis system and statistic results for a half solar cycle with Nan{\c{c}}ay Decameter Array data}",
      journal = {\aap},
         year = 2018,
        month = oct,
       volume = {618},
          eid = {A165},
        pages = {A165},
          doi = {10.1051/0004-6361/201833260},
archivePrefix = {arXiv},
       eprint = {1810.02921},
 primaryClass = {astro-ph.SR},
       adsurl = {https://ui.adsabs.harvard.edu/abs/2018A&A...618A.165Z}
}

@ARTICLE{2003ApJ...590..533R,
       author = {{Reiner}, M.~J. and {Vourlidas}, A. and {Cyr}, O.~C. St. and {Burkepile}, J.~T. and {Howard}, R.~A. and {Kaiser}, M.~L. and {Prestage}, N.~P. and {Bougeret}, J. -L.},
        title = "{Constraints on Coronal Mass Ejection Dynamics from Simultaneous Radio and White-Light Observations}",
      journal = {\apj},
         year = 2003,
        month = jun,
       volume = {590},
       number = {1},
        pages = {533-546},
          doi = {10.1086/374917},
       adsurl = {https://ui.adsabs.harvard.edu/abs/2003ApJ...590..533R}
}

@ARTICLE{2021ApJ...913...99K,
       author = {{Kouloumvakos}, Athanasios and {Rouillard}, Alexis and {Warmuth}, Alexander and {Magdalenic}, Jasmina and {Jebaraj}, Immanuel. C. and {Mann}, Gottfried and {Vainio}, Rami and {Monstein}, Christian},
        title = "{Coronal Conditions for the Occurrence of Type II Radio Bursts}",
      journal = {\apj},
         year = 2021,
        month = jun,
       volume = {913},
       number = {2},
          eid = {99},
        pages = {99},
          doi = {10.3847/1538-4357/abf435},
       adsurl = {https://ui.adsabs.harvard.edu/abs/2021ApJ...913...99K}
}

@ARTICLE{2020ApJ...893...76K,
       author = {{Kouloumvakos}, Athanasios and {Rouillard}, Alexis P. and {Share}, Gerald H. and {Plotnikov}, Illya and {Murphy}, Ronald and {Papaioannou}, Athanasios and {Wu}, Yihong},
        title = "{Evidence for a Coronal Shock Wave Origin for Relativistic Protons Producing Solar Gamma-Rays and Observed by Neutron Monitors at Earth}",
      journal = {\apj},
         year = 2020,
        month = apr,
       volume = {893},
       number = {1},
          eid = {76},
        pages = {76},
          doi = {10.3847/1538-4357/ab8227},
archivePrefix = {arXiv},
       eprint = {2004.00355},
 primaryClass = {astro-ph.SR},
       adsurl = {https://ui.adsabs.harvard.edu/abs/2020ApJ...893...76K}
}

@ARTICLE{2016ApJ...833...45R,
       author = {{Rouillard}, A.~P. and {Plotnikov}, I. and {Pinto}, R.~F. and {Tirole}, M. and {Lavarra}, M. and {Zucca}, P. and {Vainio}, R. and {Tylka}, A.~J. and {Vourlidas}, A. and {De Rosa}, M.~L. and {Linker}, J. and {Warmuth}, A. and {Mann}, G. and {Cohen}, C.~M.~S. and {Mewaldt}, R.~A.},
        title = "{Deriving the Properties of Coronal Pressure Fronts in 3D: Application to the 2012 May 17 Ground Level Enhancement}",
      journal = {\apj},
         year = 2016,
        month = dec,
       volume = {833},
       number = {1},
          eid = {45},
        pages = {45},
          doi = {10.3847/1538-4357/833/1/45},
archivePrefix = {arXiv},
       eprint = {1605.05208},
 primaryClass = {astro-ph.SR},
       adsurl = {https://ui.adsabs.harvard.edu/abs/2016ApJ...833...45R}
}

@ARTICLE{2022ApJ...926..227F,
       author = {{Frassati}, Federica and {Laurenza}, Monica and {Bemporad}, Alessandro and {West}, Matthew J. and {Mancuso}, Salvatore and {Susino}, Roberto and {Alberti}, Tommaso and {Romano}, Paolo},
        title = "{Acceleration of Solar Energetic Particles through CME-driven Shock and Streamer Interaction}",
      journal = {\apj},
         year = 2022,
        month = feb,
       volume = {926},
       number = {2},
          eid = {227},
        pages = {227},
          doi = {10.3847/1538-4357/ac460e},
       adsurl = {https://ui.adsabs.harvard.edu/abs/2022ApJ...926..227F}
}

@ARTICLE{2023ApJ...954..203L,
       author = {{Liu}, Wenlong and {Kong}, Xiangliang and {Guo}, Fan and {Zhao}, Lulu and {Feng}, Shiwei and {Yu}, Feiyu and {Jiang}, Zelong and {Chen}, Yao and {Giacalone}, Joe},
        title = "{Effects of Coronal Magnetic Field Configuration on Particle Acceleration and Release during the Ground Level Enhancement Events in Solar Cycle 24}",
      journal = {\apj},
         year = 2023,
        month = sep,
       volume = {954},
       number = {2},
          eid = {203},
        pages = {203},
          doi = {10.3847/1538-4357/ace9d2},
archivePrefix = {arXiv},
       eprint = {2307.12191},
 primaryClass = {astro-ph.SR},
       adsurl = {https://ui.adsabs.harvard.edu/abs/2023ApJ...954..203L}
}

@ARTICLE{2021GeoRL..4895138L,
       author = {{Li}, G. and {Wu}, X. and {Effenberger}, F. and {Zhao}, L. and {Lesage}, S. and {Bian}, N. and {Wang}, L.},
        title = "{Constraints on the Electron Acceleration Process in Solar Flare: A Case Study}",
      journal = {\grl},
         year = 2021,
        month = oct,
       volume = {48},
       number = {20},
          eid = {e95138},
        pages = {e95138},
          doi = {10.1029/2021GL095138},
       adsurl = {https://ui.adsabs.harvard.edu/abs/2021GeoRL..4895138L}
}

@ARTICLE{2005A&A...435.1123W,
       author = {{Warmuth}, A. and {Mann}, G.},
        title = "{A model of the Alfv{\'e}n speed in the solar corona}",
      journal = {\aap},
         year = 2005,
        month = jun,
       volume = {435},
       number = {3},
        pages = {1123-1135},
          doi = {10.1051/0004-6361:20042169},
       adsurl = {https://ui.adsabs.harvard.edu/abs/2005A&A...435.1123W}
}

@ARTICLE{2008SoPh..253..215V,
       author = {{Vr{\v{s}}nak}, Bojan and {Cliver}, Edward W.},
        title = "{Origin of Coronal Shock Waves. Invited Review}",
      journal = {\solphys},
         year = 2008,
        month = dec,
       volume = {253},
       number = {1-2},
          eid = {215},
        pages = {215},
          doi = {10.1007/s11207-008-9241-5},
       adsurl = {https://ui.adsabs.harvard.edu/abs/2008SoPh..253..215V}
}

@ARTICLE{2023ApJ...956..112K,
       author = {{Kontar}, Eduard P. and {Emslie}, A. Gordon and {Clarkson}, Daniel L. and {Chen}, Xingyao and {Chrysaphi}, Nicolina and {Azzollini}, Francesco and {Jeffrey}, Natasha L.~S. and {Gordovskyy}, Mykola},
        title = "{An Anisotropic Density Turbulence Model from the Sun to 1 au Derived from Radio Observations}",
      journal = {\apj},
         year = 2023,
        month = oct,
       volume = {956},
       number = {2},
          eid = {112},
        pages = {112},
          doi = {10.3847/1538-4357/acf6c1},
archivePrefix = {arXiv},
       eprint = {2308.05839},
 primaryClass = {astro-ph.SR},
       adsurl = {https://ui.adsabs.harvard.edu/abs/2023ApJ...956..112K}
}

@ARTICLE{2019ApJ...871..105Z,
       author = {{Zhong}, Z. and {Guo}, Y. and {Ding}, M.~D. and {Fang}, C. and {Hao}, Q.},
        title = "{Transition from Circular-ribbon to Parallel-ribbon Flares Associated with a Bifurcated Magnetic Flux Rope}",
      journal = {\apj},
         year = 2019,
        month = jan,
       volume = {871},
       number = {1},
          eid = {105},
        pages = {105},
          doi = {10.3847/1538-4357/aaf863},
archivePrefix = {arXiv},
       eprint = {1812.10223},
 primaryClass = {astro-ph.SR},
       adsurl = {https://ui.adsabs.harvard.edu/abs/2019ApJ...871..105Z}
}

@ARTICLE{2023ApJ...952...51K,
       author = {{Koval}, Artem and {Stanislavsky}, Aleksander and {Karlick{\'y}}, Marian and {Wang}, Bing and {Yerin}, Serge and {Konovalenko}, Aleksander and {B{\'a}rta}, Miroslav},
        title = "{Morphology of Solar Type II Bursts Caused by Shock Propagation through Turbulent and Inhomogeneous Coronal Plasma}",
      journal = {\apj},
         year = 2023,
        month = jul,
       volume = {952},
       number = {1},
          eid = {51},
        pages = {51},
          doi = {10.3847/1538-4357/acdbcc},
       adsurl = {https://ui.adsabs.harvard.edu/abs/2023ApJ...952...51K}
}

@ARTICLE{2012ApJ...754....9G,
       author = {{Glesener}, Lindsay and {Krucker}, S{\"a}m and {Lin}, R.~P.},
        title = "{Hard X-Ray Observations of a Jet and Accelerated Electrons in the Corona}",
      journal = {\apj},
         year = 2012,
        month = jul,
       volume = {754},
       number = {1},
          eid = {9},
        pages = {9},
          doi = {10.1088/0004-637X/754/1/9},
       adsurl = {https://ui.adsabs.harvard.edu/abs/2012ApJ...754....9G}
}

@ARTICLE{2018ApJ...866...62C,
       author = {{Chen}, Bin and {Yu}, Sijie and {Battaglia}, Marina and {Farid}, Samaiyah and {Savcheva}, Antonia and {Reeves}, Katharine K. and {Krucker}, S{\"a}m and {Bastian}, T.~S. and {Guo}, Fan and {Tassev}, Svetlin},
        title = "{Magnetic Reconnection Null Points as the Origin of Semirelativistic Electron Beams in a Solar Jet}",
      journal = {\apj},
         year = 2018,
        month = oct,
       volume = {866},
       number = {1},
          eid = {62},
        pages = {62},
          doi = {10.3847/1538-4357/aadb89},
archivePrefix = {arXiv},
       eprint = {1808.05951},
 primaryClass = {astro-ph.SR},
       adsurl = {https://ui.adsabs.harvard.edu/abs/2018ApJ...866...62C}
}

@ARTICLE{2024RvMPP...8....7Z,
       author = {{Zhang}, Qingmin},
        title = "{Circular-ribbon flares and the related activities}",
      journal = {Reviews of Modern Plasma Physics},
         year = 2024,
        month = mar,
       volume = {8},
       number = {1},
          eid = {7},
        pages = {7},
          doi = {10.1007/s41614-024-00144-9},
archivePrefix = {arXiv},
       eprint = {2401.16101},
 primaryClass = {astro-ph.SR},
       adsurl = {https://ui.adsabs.harvard.edu/abs/2024RvMPP...8....7Z}
}

@ARTICLE{2022ApJS..260...19Z,
       author = {{Zhang}, Yanjie and {Zhang}, Qingmin and {Song}, Dechao and {Li}, Shuting and {Dai}, Jun and {Xu}, Zhe and {Ji}, Haisheng},
        title = "{Statistical Analysis of Circular-ribbon Flares}",
      journal = {\apjs},
         year = 2022,
        month = may,
       volume = {260},
       number = {1},
          eid = {19},
        pages = {19},
          doi = {10.3847/1538-4365/ac5f4c},
archivePrefix = {arXiv},
       eprint = {2203.12819},
 primaryClass = {astro-ph.SR},
       adsurl = {https://ui.adsabs.harvard.edu/abs/2022ApJS..260...19Z}
}

@ARTICLE{2017SoPh..292..194L,
       author = {{Lv}, M.~S. and {Chen}, Y. and {Li}, C.~Y. and {Zimovets}, I. and {Du}, G.~H. and {Wang}, B. and {Feng}, S.~W. and {Ma}, S.~L.},
        title = "{Sources of the Multi-Lane Type II Solar Radio Burst on 5 November 2014}",
      journal = {\solphys},
         year = 2017,
        month = dec,
       volume = {292},
       number = {12},
          eid = {194},
        pages = {194},
          doi = {10.1007/s11207-017-1218-9},
       adsurl = {https://ui.adsabs.harvard.edu/abs/2017SoPh..292..194L}
}

@ARTICLE{2024SoPh..299...63V,
       author = {{Vasanth}, V.},
        title = "{Coronal Signatures of Flare Generated Fast-Mode Wave at EUV and Radio Wavelengths}",
      journal = {\solphys},
         year = 2024,
        month = may,
       volume = {299},
       number = {5},
          eid = {63},
        pages = {63},
          doi = {10.1007/s11207-024-02293-z},
archivePrefix = {arXiv},
       eprint = {2404.00135},
 primaryClass = {astro-ph.SR},
       adsurl = {https://ui.adsabs.harvard.edu/abs/2024SoPh..299...63V}
}

@ARTICLE{2024A&A...689A.345K,
       author = {{Koval}, Artem and {Karlick{\'y}}, Marian and {Brazhenko}, Anatolii and {Stanislavsky}, Aleksander and {Frantsuzenko}, Anatolii and {Vandas}, Marek and {Konovalenko}, Aleksander and {B{\'a}rta}, Miroslav and {Bubnov}, Ihor and {Miteva}, Rositsa and {Yerin}, Serge},
        title = "{Spectral cleaving in solar type II radio bursts: Observations and interpretation}",
      journal = {\aap},
         year = 2024,
        month = sep,
       volume = {689},
          eid = {A345},
        pages = {A345},
          doi = {10.1051/0004-6361/202451010},
       adsurl = {https://ui.adsabs.harvard.edu/abs/2024A&A...689A.345K}
}

@ARTICLE{2012A&A...547A...6Z,
       author = {{Zimovets}, I. and {Vilmer}, N. and {Chian}, A.~C. -L. and {Sharykin}, I. and {Struminsky}, A.},
        title = "{Spatially resolved observations of a split-band coronal type II radio burst}",
      journal = {\aap},
         year = 2012,
        month = nov,
       volume = {547},
          eid = {A6},
        pages = {A6},
          doi = {10.1051/0004-6361/201219454},
archivePrefix = {arXiv},
       eprint = {1208.5267},
 primaryClass = {astro-ph.SR},
       adsurl = {https://ui.adsabs.harvard.edu/abs/2012A&A...547A...6Z}
}

@ARTICLE{2016ApJ...827L...9F,
       author = {{Feng}, S.~W. and {Chen}, Y. and {Song}, H.~Q. and {Wang}, B. and {Kong}, X.~L.},
        title = "{An Imaging Study of a Complex Solar Coronal Radio Eruption}",
      journal = {\apjl},
         year = 2016,
        month = aug,
       volume = {827},
       number = {1},
          eid = {L9},
        pages = {L9},
          doi = {10.3847/2041-8205/827/1/L9},
archivePrefix = {arXiv},
       eprint = {1608.00073},
 primaryClass = {astro-ph.SR},
       adsurl = {https://ui.adsabs.harvard.edu/abs/2016ApJ...827L...9F}
}

@ARTICLE{2024ApJS..272...21C,
       author = {{Chang}, ShuWang and {Wang}, Bing and {Lu}, Guang and {Shen}, YuPeng and {Bai}, Yu and {Shang}, ZiQian and {Zhang}, Lei and {Wu}, Zhao and {Su}, YanRui and {Chen}, Yao and {Yan}, FaBao},
        title = "{Development of a 90{\textendash}600 MHz Meter-wave Solar Radio Spectrometer}",
      journal = {\apjs},
         year = 2024,
        month = may,
       volume = {272},
       number = {1},
          eid = {21},
        pages = {21},
          doi = {10.3847/1538-4365/ad3de7},
       adsurl = {https://ui.adsabs.harvard.edu/abs/2024ApJS..272...21C}
}

@ARTICLE{2024ApJ...964..108F,
       author = {{Feng}, S.~W. and {Xie}, H.~X. and {Misawa}, H.},
        title = "{Solar Type J Radio Bursts and the Associated Coronal Loop}",
      journal = {\apj},
         year = 2024,
        month = apr,
       volume = {964},
       number = {2},
          eid = {108},
        pages = {108},
          doi = {10.3847/1538-4357/ad267f},
       adsurl = {https://ui.adsabs.harvard.edu/abs/2024ApJ...964..108F}
}

@ARTICLE{2023ApJ...953..171H,
       author = {{Hou}, Zhenyong and {Tian}, Hui and {Su}, Wei and {Madjarska}, Maria S. and {Chen}, Hechao and {Zheng}, Ruisheng and {Bai}, Xianyong and {Deng}, Yuanyong},
        title = "{A Type II Radio Burst Driven by a Blowout Jet on the Sun}",
      journal = {\apj},
         year = 2023,
        month = aug,
       volume = {953},
       number = {2},
          eid = {171},
        pages = {171},
          doi = {10.3847/1538-4357/ace31b},
archivePrefix = {arXiv},
       eprint = {2306.16725},
 primaryClass = {astro-ph.SR},
       adsurl = {https://ui.adsabs.harvard.edu/abs/2023ApJ...953..171H}
}

@ARTICLE{2023A&A...675A..98M,
       author = {{Morosan}, D.~E. and {Pomoell}, J. and {Kumari}, A. and {Kilpua}, E.~K.~J. and {Vainio}, R.},
        title = "{A type II solar radio burst without a coronal mass ejection}",
      journal = {\aap},
         year = 2023,
        month = jul,
       volume = {675},
          eid = {A98},
        pages = {A98},
          doi = {10.1051/0004-6361/202245515},
archivePrefix = {arXiv},
       eprint = {2305.11545},
 primaryClass = {astro-ph.SR},
       adsurl = {https://ui.adsabs.harvard.edu/abs/2023A&A...675A..98M}
}

@ARTICLE{2023A&A...675A.102K,
       author = {{Kumari}, A. and {Morosan}, D.~E. and {Kilpua}, E.~K.~J. and {Daei}, F.},
        title = "{Type II radio bursts and their association with coronal mass ejections in solar cycles 23 and 24}",
      journal = {\aap},
         year = 2023,
        month = jul,
       volume = {675},
          eid = {A102},
        pages = {A102},
          doi = {10.1051/0004-6361/202244015},
archivePrefix = {arXiv},
       eprint = {2305.18992},
 primaryClass = {astro-ph.SR},
       adsurl = {https://ui.adsabs.harvard.edu/abs/2023A&A...675A.102K}
}

@ARTICLE{2023A&A...669A..28M,
       author = {{Mancuso}, S. and {Barghini}, D. and {Bemporad}, A. and {Telloni}, D. and {Gardiol}, D. and {Frassati}, F. and {Bizzarri}, I. and {Taricco}, C.},
        title = "{Three-dimensional reconstruction of type U radio bursts: a novel remote sensing approach for coronal loops}",
      journal = {\aap},
         year = 2023,
        month = jan,
       volume = {669},
          eid = {A28},
        pages = {A28},
          doi = {10.1051/0004-6361/202243841},
archivePrefix = {arXiv},
       eprint = {2212.02147},
 primaryClass = {astro-ph.SR},
       adsurl = {https://ui.adsabs.harvard.edu/abs/2023A&A...669A..28M}
}

@ARTICLE{2020FrASS...7...56R,
       author = {{Reid}, Hamish A.~S.},
        title = "{A review of recent type III imaging spectroscopy}",
      journal = {Frontiers in Astronomy and Space Sciences},
         year = 2020,
        month = sep,
       volume = {7},
          eid = {56},
        pages = {56},
          doi = {10.3389/fspas.2020.00056},
       adsurl = {https://ui.adsabs.harvard.edu/abs/2020FrASS...7...56R}
}

@ARTICLE{2019RAA....19..163S,
       author = {{Su}, Yang and {Liu}, Wei and {Li}, You-Ping and {Zhang}, Zhe and {Hurford}, Gordon J. and {Chen}, Wei and {Huang}, Yu and {Li}, Zhen-Tong and {Jiang}, Xian-Kai and {Wang}, Hao-Xiang and {Xia}, Fan-Xiao-Yu and {Chen}, Chang-Xue and {Yu}, Wen-Hui and {Yu}, Fu and {Wu}, Jian and {Gan}, Wei-Qun},
        title = "{Simulations and software development for the Hard X-ray Imager onboard ASO-S}",
      journal = {Research in Astronomy and Astrophysics},
         year = 2019,
        month = nov,
       volume = {19},
       number = {11},
          eid = {163},
        pages = {163},
          doi = {10.1088/1674-4527/19/11/163},
       adsurl = {https://ui.adsabs.harvard.edu/abs/2019RAA....19..163S}
}

@ARTICLE{2019RAA....19..156G,
       author = {{Gan}, Wei-Qun and {Zhu}, Cheng and {Deng}, Yuan-Yong and {Li}, Hui and {Su}, Yang and {Zhang}, Hai-Ying and {Chen}, Bo and {Zhang}, Zhe and {Wu}, Jian and {Deng}, Lei and {Huang}, Yu and {Yang}, Jian-Feng and {Cui}, Ji-Jun and {Chang}, Jin and {Wang}, Chi and {Wu}, Ji and {Yin}, Zeng-Shan and {Chen}, Wen and {Fang}, Cheng and {Yan}, Yi-Hua and {Lin}, Jun and {Xiong}, Wei-Ming and {Chen}, Bin and {Bao}, Hai-Chao and {Cao}, Cai-Xia and {Bai}, Yan-Ping and {Wang}, Tao and {Chen}, Bing-Long and {Li}, Xin-Yu and {Zhang}, Ye and {Feng}, Li and {Su}, Jiang-Tao and {Li}, Ying and {Chen}, Wei and {Li}, You-Ping and {Su}, Ying-Na and {Wu}, Hai-Yan and {Gu}, Mei and {Huang}, Lei and {Tang}, Xue-Jun},
        title = "{Advanced Space-based Solar Observatory (ASO-S): an overview}",
      journal = {Research in Astronomy and Astrophysics},
         year = 2019,
        month = nov,
       volume = {19},
       number = {11},
          eid = {156},
        pages = {156},
          doi = {10.1088/1674-4527/19/11/156},
       adsurl = {https://ui.adsabs.harvard.edu/abs/2019RAA....19..156G}
}

@ARTICLE{2019NatAs...3..452M,
       author = {{Morosan}, Diana E. and {Carley}, Eoin P. and {Hayes}, Laura A. and {Murray}, Sophie A. and {Zucca}, Pietro and {Fallows}, Richard A. and {McCauley}, Joe and {Kilpua}, Emilia K.~J. and {Mann}, Gottfried and {Vocks}, Christian and {Gallagher}, Peter T.},
        title = "{Multiple regions of shock-accelerated particles during a solar coronal mass ejection}",
      journal = {Nature Astronomy},
         year = 2019,
        month = feb,
       volume = {3},
        pages = {452-461},
          doi = {10.1038/s41550-019-0689-z},
archivePrefix = {arXiv},
       eprint = {1908.11743},
 primaryClass = {astro-ph.SR},
       adsurl = {https://ui.adsabs.harvard.edu/abs/2019NatAs...3..452M}
}

@ARTICLE{2017A&A...606A.141R,
       author = {{Reid}, Hamish A.~S. and {Kontar}, Eduard P.},
        title = "{Imaging spectroscopy of type U and J solar radio bursts with LOFAR}",
      journal = {\aap},
         year = 2017,
        month = oct,
       volume = {606},
          eid = {A141},
        pages = {A141},
          doi = {10.1051/0004-6361/201730701},
archivePrefix = {arXiv},
       eprint = {1706.07410},
 primaryClass = {astro-ph.SR},
       adsurl = {https://ui.adsabs.harvard.edu/abs/2017A&A...606A.141R}
}

@ARTICLE{2016ApJ...830...37K,
       author = {{Kong}, Xiangliang and {Chen}, Yao and {Feng}, Shiwei and {Du}, Guohui and {Li}, Chuanyang and {Koval}, Artem and {Vasanth}, V. and {Wang}, Bing and {Guo}, Fan and {Li}, Gang},
        title = "{Observation of a Metric Type N Solar Radio Burst}",
      journal = {\apj},
         year = 2016,
        month = oct,
       volume = {830},
       number = {1},
          eid = {37},
        pages = {37},
          doi = {10.3847/0004-637X/830/1/37},
archivePrefix = {arXiv},
       eprint = {1608.00093},
 primaryClass = {astro-ph.SR},
       adsurl = {https://ui.adsabs.harvard.edu/abs/2016ApJ...830...37K}
}

@ARTICLE{2016ApJ...828...28K,
       author = {{Kumar}, Pankaj and {Innes}, D.~E. and {Cho}, Kyung-Suk},
        title = "{Flare-generated Shock Wave Propagation through Solar Coronal Arcade Loops and an Associated Type II Radio Burst}",
      journal = {\apj},
         year = 2016,
        month = sep,
       volume = {828},
       number = {1},
          eid = {28},
        pages = {28},
          doi = {10.3847/0004-637X/828/1/28},
archivePrefix = {arXiv},
       eprint = {1606.05056},
 primaryClass = {astro-ph.SR},
       adsurl = {https://ui.adsabs.harvard.edu/abs/2016ApJ...828...28K}
}

@ARTICLE{2015ApJ...804...88S,
       author = {{Su}, W. and {Cheng}, X. and {Ding}, M.~D. and {Chen}, P.~F. and {Sun}, J.~Q.},
        title = "{A Type II Radio Burst without a Coronal Mass Ejection}",
      journal = {\apj},
         year = 2015,
        month = may,
       volume = {804},
       number = {2},
          eid = {88},
        pages = {88},
          doi = {10.1088/0004-637X/804/2/88},
archivePrefix = {arXiv},
       eprint = {1503.00861},
 primaryClass = {astro-ph.SR},
       adsurl = {https://ui.adsabs.harvard.edu/abs/2015ApJ...804...88S}
}

@ARTICLE{2015ApJ...798...81K,
       author = {{Kong}, Xiangliang and {Chen}, Yao and {Guo}, Fan and {Feng}, Shiwei and {Wang}, Bing and {Du}, Guohui and {Li}, Gang},
        title = "{The Possible Role of Coronal Streamers as Magnetically Closed Structures in Shock-induced Energetic Electrons and Metric Type II Radio Bursts}",
      journal = {\apj},
         year = 2015,
        month = jan,
       volume = {798},
       number = {2},
          eid = {81},
        pages = {81},
          doi = {10.1088/0004-637X/798/2/81},
archivePrefix = {arXiv},
       eprint = {1410.7994},
 primaryClass = {astro-ph.SR},
       adsurl = {https://ui.adsabs.harvard.edu/abs/2015ApJ...798...81K}
}

@ARTICLE{2012ApJ...746..152M,
       author = {{Magdaleni{\'c}}, J. and {Marqu{\'e}}, C. and {Zhukov}, A.~N. and {Vr{\v{s}}nak}, B. and {Veronig}, A.},
        title = "{Flare-generated Type II Burst without Associated Coronal Mass Ejection}",
      journal = {\apj},
         year = 2012,
        month = feb,
       volume = {746},
       number = {2},
          eid = {152},
        pages = {152},
          doi = {10.1088/0004-637X/746/2/152},
       adsurl = {https://ui.adsabs.harvard.edu/abs/2012ApJ...746..152M}
}

@ARTICLE{2012SoPh..275...17L,
       author = {{Lemen}, James R. and {Title}, Alan M. and {Akin}, David J. and {Boerner}, Paul F. and {Chou}, Catherine and {Drake}, Jerry F. and {Duncan}, Dexter W. and {Edwards}, Christopher G. and {Friedlaender}, Frank M. and {Heyman}, Gary F. and {Hurlburt}, Neal E. and {Katz}, Noah L. and {Kushner}, Gary D. and {Levay}, Michael and {Lindgren}, Russell W. and {Mathur}, Dnyanesh P. and {McFeaters}, Edward L. and {Mitchell}, Sarah and {Rehse}, Roger A. and {Schrijver}, Carolus J. and {Springer}, Larry A. and {Stern}, Robert A. and {Tarbell}, Theodore D. and {Wuelser}, Jean-Pierre and {Wolfson}, C. Jacob and {Yanari}, Carl and {Bookbinder}, Jay A. and {Cheimets}, Peter N. and {Caldwell}, David and {Deluca}, Edward E. and {Gates}, Richard and {Golub}, Leon and {Park}, Sang and {Podgorski}, William A. and {Bush}, Rock I. and {Scherrer}, Philip H. and {Gummin}, Mark A. and {Smith}, Peter and {Auker}, Gary and {Jerram}, Paul and {Pool}, Peter and {Soufli}, Regina and {Windt}, David L. and {Beardsley}, Sarah and {Clapp}, Matthew and {Lang}, James and {Waltham}, Nicholas},
        title = "{The Atmospheric Imaging Assembly (AIA) on the Solar Dynamics Observatory (SDO)}",
      journal = {\solphys},
         year = 2012,
        month = jan,
       volume = {275},
       number = {1-2},
        pages = {17-40},
          doi = {10.1007/s11207-011-9776-8},
       adsurl = {https://ui.adsabs.harvard.edu/abs/2012SoPh..275...17L}
}

@ARTICLE{2012SoPh..275....3P,
       author = {{Pesnell}, W. Dean and {Thompson}, B.~J. and {Chamberlin}, P.~C.},
        title = "{The Solar Dynamics Observatory (SDO)}",
      journal = {\solphys},
         year = 2012,
        month = jan,
       volume = {275},
       number = {1-2},
        pages = {3-15},
          doi = {10.1007/s11207-011-9841-3},
       adsurl = {https://ui.adsabs.harvard.edu/abs/2012SoPh..275....3P}
}

@ARTICLE{2011A&A...531A..31N,
       author = {{Nindos}, A. and {Alissandrakis}, C.~E. and {Hillaris}, A. and {Preka-Papadema}, P.},
        title = "{On the relationship of shock waves to flares and coronal mass ejections}",
      journal = {\aap},
         year = 2011,
        month = jul,
       volume = {531},
          eid = {A31},
        pages = {A31},
          doi = {10.1051/0004-6361/201116799},
archivePrefix = {arXiv},
       eprint = {1105.1268},
 primaryClass = {astro-ph.SR},
       adsurl = {https://ui.adsabs.harvard.edu/abs/2011A&A...531A..31N}
}

@ARTICLE{2008A&ARv..16....1P,
       author = {{Pick}, Monique and {Vilmer}, Nicole},
        title = "{Sixty-five years of solar radioastronomy: flares, coronal mass ejections and Sun Earth connection}",
      journal = {\aapr},
         year = 2008,
        month = oct,
       volume = {16},
        pages = {1-153},
          doi = {10.1007/s00159-008-0013-x},
       adsurl = {https://ui.adsabs.harvard.edu/abs/2008A&ARv..16....1P}
}

@INCOLLECTION{1997LNP...483..192K,
       author = {{Kerdraon}, Alain and {Delouis}, Jean-Marc},
        title = "{The Nan{\c{c}}ay Radioheliograph}",
    booktitle = {Coronal Physics from Radio and Space Observations},
         year = 1997,
       editor = {{Trottet}, Gerard},
       volume = {483},
        pages = {192},
          doi = {10.1007/BFb0106458},
       adsurl = {https://ui.adsabs.harvard.edu/abs/1997LNP...483..192K}
}

@ARTICLE{1995SoPh..162..357B,
       author = {{Brueckner}, G.~E. and {Howard}, R.~A. and {Koomen}, M.~J. and {Korendyke}, C.~M. and {Michels}, D.~J. and {Moses}, J.~D. and {Socker}, D.~G. and {Dere}, K.~P. and {Lamy}, P.~L. and {Llebaria}, A. and {Bout}, M.~V. and {Schwenn}, R. and {Simnett}, G.~M. and {Bedford}, D.~K. and {Eyles}, C.~J.},
        title = "{The Large Angle Spectroscopic Coronagraph (LASCO)}",
      journal = {\solphys},
         year = 1995,
        month = dec,
       volume = {162},
       number = {1-2},
        pages = {357-402},
          doi = {10.1007/BF00733434},
       adsurl = {https://ui.adsabs.harvard.edu/abs/1995SoPh..162..357B}
}

@ARTICLE{1987ApJ...319..503C,
       author = {{Caroubalos}, C. and {Poquerusse}, M. and {Bougeret}, J. -L. and {Crepel}, R.},
        title = "{Radio Evidence for a Magnetic Mirror Effect on Beams of Subrelativistic Electrons in the Solar Corona}",
      journal = {\apj},
         year = 1987,
        month = aug,
       volume = {319},
        pages = {503},
          doi = {10.1086/165473},
       adsurl = {https://ui.adsabs.harvard.edu/abs/1987ApJ...319..503C}
}

@INCOLLECTION{1985srph.book..333N,
       author = {{Nelson}, G.~J. and {Melrose}, D.~B.},
        title = "{Type II bursts.}",
    booktitle = {Solar Radiophysics: Studies of Emission from the Sun at Metre Wavelengths},
         year = 1985,
       editor = {{McLean}, D.~J. and {Labrum}, N.~R.},
        pages = {333-359},
       adsurl = {https://ui.adsabs.harvard.edu/abs/1985srph.book..333N}
}

@ARTICLE{1983ApJ...267..837H,
       author = {{Holman}, G.~D. and {Pesses}, M.~E.},
        title = "{Solar type II radio emission and the shock drift acceleration of electrons}",
      journal = {\apj},
         year = 1983,
        month = apr,
       volume = {267},
        pages = {837-843},
          doi = {10.1086/160918},
       adsurl = {https://ui.adsabs.harvard.edu/abs/1983ApJ...267..837H}
}

@ARTICLE{1975ApL....16...23S,
       author = {{Smerd}, S.~F. and {Sheridan}, K.~V. and {Stewart}, R.~T.},
        title = "{Split-Band Structure in Type II Radio Bursts from the Sun}",
      journal = {\aplett},
         year = 1975,
        month = feb,
       volume = {16},
        pages = {23},
       adsurl = {https://ui.adsabs.harvard.edu/abs/1975ApL....16...23S}
}

@ARTICLE{1958AZh....35..694G,
       author = {{Ginzburg}, V.~L. and {Zhelezniakov}, V.~V.},
        title = "{On the Possible Mechanisms of Sporadic Solar Radio Emission (Radiation in an Isotropic Plasma)}",
      journal = {\azh},
         year = 1958,
        month = jan,
       volume = {35},
        pages = {694},
       adsurl = {https://ui.adsabs.harvard.edu/abs/1958AZh....35..694G}
}

@ARTICLE{2009ApJ...702..791M,
       author = {{Meegan}, Charles and {Lichti}, Giselher and {Bhat}, P.~N. and {Bissaldi}, Elisabetta and {Briggs}, Michael S. and {Connaughton}, Valerie and {Diehl}, Roland and {Fishman}, Gerald and {Greiner}, Jochen and {Hoover}, Andrew S. and {van der Horst}, Alexander J. and {von Kienlin}, Andreas and {Kippen}, R. Marc and {Kouveliotou}, Chryssa and {McBreen}, Sheila and {Paciesas}, W.~S. and {Preece}, Robert and {Steinle}, Helmut and {Wallace}, Mark S. and {Wilson}, Robert B. and {Wilson-Hodge}, Colleen},
        title = "{The Fermi Gamma-ray Burst Monitor}",
      journal = {\apj},
         year = 2009,
        month = sep,
       volume = {702},
       number = {1},
        pages = {791-804},
          doi = {10.1088/0004-637X/702/1/791},
archivePrefix = {arXiv},
       eprint = {0908.0450},
 primaryClass = {astro-ph.IM},
       adsurl = {https://ui.adsabs.harvard.edu/abs/2009ApJ...702..791M}
}

@ARTICLE{2022ApJ...926L..39D,
       author = {{Duan}, Yadan and {Shen}, Yuandeng and {Zhou}, Xinping and {Tang}, Zehao and {Zhou}, Chengrui and {Tan}, Song},
        title = "{Homologous Accelerated Electron Beams, a Quasiperiodic Fast-propagating Wave, and a Coronal Mass Ejection Observed in One Fan-spine Jet}",
      journal = {\apjl},
         year = 2022,
        month = feb,
       volume = {926},
       number = {2},
          eid = {L39},
        pages = {L39},
          doi = {10.3847/2041-8213/ac4df2},
archivePrefix = {arXiv},
       eprint = {2201.08982},
 primaryClass = {astro-ph.SR},
       adsurl = {https://ui.adsabs.harvard.edu/abs/2022ApJ...926L..39D}
}

@ARTICLE{2024ApJ...968..110D,
       author = {{Duan}, Yadan and {Shen}, Yuandeng and {Tang}, Zehao and {Zhou}, Chenrui and {Tan}, Song},
        title = "{On the Determining Physical Factor of Jet-related Coronal Mass Ejections' Morphology in the High Corona}",
      journal = {\apj},
         year = 2024,
        month = jun,
       volume = {968},
       number = {2},
          eid = {110},
        pages = {110},
          doi = {10.3847/1538-4357/ad445c},
archivePrefix = {arXiv},
       eprint = {2404.19179},
 primaryClass = {astro-ph.SR},
       adsurl = {https://ui.adsabs.harvard.edu/abs/2024ApJ...968..110D}
}

@ARTICLE{2015ApJ...812...52D,
       author = {{Du}, Guohui and {Kong}, Xiangliang and {Chen}, Yao and {Feng}, Shiwei and {Wang}, Bing and {Li}, Gang},
        title = "{An Observational Revisit of Band-split Solar Type-II Radio Bursts}",
      journal = {\apj},
         year = 2015,
        month = oct,
       volume = {812},
       number = {1},
          eid = {52},
        pages = {52},
          doi = {10.1088/0004-637X/812/1/52},
archivePrefix = {arXiv},
       eprint = {1509.03832},
 primaryClass = {astro-ph.SR},
       adsurl = {https://ui.adsabs.harvard.edu/abs/2015ApJ...812...52D}
}

@ARTICLE{2014ApJ...793L..39D,
       author = {{Du}, Guohui and {Chen}, Yao and {Lv}, Maoshui and {Kong}, Xiangliang and {Feng}, Shiwei and {Guo}, Fan and {Li}, Gang},
        title = "{Temporal Spectral Shift and Polarization of a Band-splitting Solar Type II Radio Burst}",
      journal = {\apjl},
         year = 2014,
        month = oct,
       volume = {793},
       number = {2},
          eid = {L39},
        pages = {L39},
          doi = {10.1088/2041-8205/793/2/L39},
       adsurl = {https://ui.adsabs.harvard.edu/abs/2014ApJ...793L..39D}
}

@ARTICLE{2001A&A...377..321V,
       author = {{Vr{\v{s}}nak}, B. and {Aurass}, H. and {Magdaleni{\'c}}, J. and {Gopalswamy}, N.},
        title = "{Band-splitting of coronal and interplanetary type II bursts. I. Basic properties}",
      journal = {\aap},
         year = 2001,
        month = oct,
       volume = {377},
        pages = {321-329},
          doi = {10.1051/0004-6361:20011067},
       adsurl = {https://ui.adsabs.harvard.edu/abs/2001A&A...377..321V}
}

@ARTICLE{2008SSRv..136...67H,
       author = {{Howard}, R.~A. and {Moses}, J.~D. and {Vourlidas}, A. and {Newmark}, J.~S. and {Socker}, D.~G. and {Plunkett}, S.~P. and {Korendyke}, C.~M. and {Cook}, J.~W. and {Hurley}, A. and {Davila}, J.~M. and {Thompson}, W.~T. and {St Cyr}, O.~C. and {Mentzell}, E. and {Mehalick}, K. and {Lemen}, J.~R. and {Wuelser}, J.~P. and {Duncan}, D.~W. and {Tarbell}, T.~D. and {Wolfson}, C.~J. and {Moore}, A. and {Harrison}, R.~A. and {Waltham}, N.~R. and {Lang}, J. and {Davis}, C.~J. and {Eyles}, C.~J. and {Mapson-Menard}, H. and {Simnett}, G.~M. and {Halain}, J.~P. and {Defise}, J.~M. and {Mazy}, E. and {Rochus}, P. and {Mercier}, R. and {Ravet}, M.~F. and {Delmotte}, F. and {Auchere}, F. and {Delaboudiniere}, J.~P. and {Bothmer}, V. and {Deutsch}, W. and {Wang}, D. and {Rich}, N. and {Cooper}, S. and {Stephens}, V. and {Maahs}, G. and {Baugh}, R. and {McMullin}, D. and {Carter}, T.},
        title = "{Sun Earth Connection Coronal and Heliospheric Investigation (SECCHI)}",
      journal = {\ssr},
         year = 2008,
        month = apr,
       volume = {136},
       number = {1-4},
        pages = {67-115},
          doi = {10.1007/s11214-008-9341-4},
       adsurl = {https://ui.adsabs.harvard.edu/abs/2008SSRv..136...67H}
}

@ARTICLE{2021ApJ...909....2M,
       author = {{Maguire}, Ciara A. and {Carley}, Eoin P. and {Zucca}, Pietro and {Vilmer}, Nicole and {Gallagher}, Peter T.},
        title = "{LOFAR Observations of a Jet-driven Piston Shock in the Low Solar Corona}",
      journal = {\apj},
         year = 2021,
        month = mar,
       volume = {909},
       number = {1},
          eid = {2},
        pages = {2},
          doi = {10.3847/1538-4357/abda51},
archivePrefix = {arXiv},
       eprint = {2101.05569},
 primaryClass = {astro-ph.SR},
       adsurl = {https://ui.adsabs.harvard.edu/abs/2021ApJ...909....2M}
}

@ARTICLE{Su2024HXI,
       author = {{Su}, Yang and {Zhang}, Zhe and {Chen}, Wei and {Chen}, Dengyi and {Yu}, Fu and {Hu}, Yiming and {Zhang}, Yan and {Xia}, Fanxiaoyu and {Chen}, Changxue and {Li}, Zhentong and {Jiang}, Xiankai and {Huang}, Yu and {Zhang}, Yongqiang and {Liu}, Wei and {Ma}, Tao and {Li}, Dong and {Yu}, Wenhui and {Li}, Youping and {Cai}, Mingsheng and {Guo}, Jianhua and {Huang}, Yongyi and {Wang}, Haoxiang and {Liang}, Yaoming and {Ma}, Miao and {Wang}, Jianping and {Zhu}, Shanshan and {Tao}, Jinyou and {Yu}, Jirui and {Yang}, Jianfeng and {Wu}, Jian and {Gan}, Weiqun},
        title = "{The Tests and Calibrations of the Hard X-ray Imager Aboard ASO-S}",
      journal = {\solphys},
         year = 2024,
        month = oct,
       volume = {299},
       number = {10},
          eid = {153},
        pages = {153},
          doi = {10.1007/s11207-024-02392-x},
       adsurl = {https://ui.adsabs.harvard.edu/abs/2024SoPh..299..153S}
}

@ARTICLE{2019ApJ...884..122K,
       author = {{Kontar}, Eduard P. and {Chen}, Xingyao and {Chrysaphi}, Nicolina and {Jeffrey}, Natasha L.~S. and {Emslie}, A. Gordon and {Krupar}, Vratislav and {Maksimovic}, Milan and {Gordovskyy}, Mykola and {Browning}, Philippa K.},
        title = "{Anisotropic Radio-wave Scattering and the Interpretation of Solar Radio Emission Observations}",
      journal = {\apj},
         year = 2019,
        month = oct,
       volume = {884},
       number = {2},
          eid = {122},
        pages = {122},
          doi = {10.3847/1538-4357/ab40bb},
archivePrefix = {arXiv},
       eprint = {1909.00340},
 primaryClass = {astro-ph.SR},
       adsurl = {https://ui.adsabs.harvard.edu/abs/2019ApJ...884..122K}
}

@ARTICLE{2014ApJ...783...11A,
       author = {{Adams}, Mitzi and {Sterling}, Alphonse C. and {Moore}, Ronald L. and {Gary}, G. Allen},
        title = "{A Small-scale Eruption Leading to a Blowout Macrospicule Jet in an On-disk Coronal Hole}",
      journal = {\apj},
         year = 2014,
        month = mar,
       volume = {783},
       number = {1},
          eid = {11},
        pages = {11},
          doi = {10.1088/0004-637X/783/1/11},
       adsurl = {https://ui.adsabs.harvard.edu/abs/2014ApJ...783...11A}
}

@ARTICLE{2015Natur.523..437S,
       author = {{Sterling}, Alphonse C. and {Moore}, Ronald L. and {Falconer}, David A. and {Adams}, Mitzi},
        title = "{Small-scale filament eruptions as the driver of X-ray jets in solar coronal holes}",
      journal = {\nat},
         year = 2015,
        month = jul,
       volume = {523},
       number = {7561},
        pages = {437-440},
          doi = {10.1038/nature14556},
archivePrefix = {arXiv},
       eprint = {1705.03373},
 primaryClass = {astro-ph.SR},
       adsurl = {https://ui.adsabs.harvard.edu/abs/2015Natur.523..437S}
}

@ARTICLE{2016ApJ...832L...7P,
       author = {{Panesar}, Navdeep K. and {Sterling}, Alphonse C. and {Moore}, Ronald L. and {Chakrapani}, Prithi},
        title = "{Magnetic Flux Cancelation as the Trigger of Solar Quiet-region Coronal Jets}",
      journal = {\apjl},
         year = 2016,
        month = nov,
       volume = {832},
       number = {1},
          eid = {L7},
        pages = {L7},
          doi = {10.3847/2041-8205/832/1/L7},
archivePrefix = {arXiv},
       eprint = {1610.08540},
 primaryClass = {astro-ph.SR},
       adsurl = {https://ui.adsabs.harvard.edu/abs/2016ApJ...832L...7P}
}

@ARTICLE{2019ApJ...882...16M,
       author = {{McGlasson}, Riley A. and {Panesar}, Navdeep K. and {Sterling}, Alphonse C. and {Moore}, Ronald L.},
        title = "{Magnetic Flux Cancellation as the Trigger Mechanism of Solar Coronal Jets}",
      journal = {\apj},
         year = 2019,
        month = sep,
       volume = {882},
       number = {1},
          eid = {16},
        pages = {16},
          doi = {10.3847/1538-4357/ab2fe3},
archivePrefix = {arXiv},
       eprint = {1906.06452},
 primaryClass = {astro-ph.SR},
       adsurl = {https://ui.adsabs.harvard.edu/abs/2019ApJ...882...16M}
}

@ARTICLE{2025ApJ...987L...4C,
       author = {{Chen}, Changxue and {Su}, Yang and {Chen}, Wei and {Li}, Jingwei and {Yu}, Fu and {Gan}, Weiqun},
        title = "{Intense Hard X-Ray Emissions in C-class Flares: A Statistical Study with ASO-S/HXI Data}",
      journal = {\apjl},
         year = 2025,
        month = jul,
       volume = {987},
       number = {1},
          eid = {L4},
        pages = {L4},
          doi = {10.3847/2041-8213/addf2f},
archivePrefix = {arXiv},
       eprint = {2506.12294},
 primaryClass = {astro-ph.SR},
       adsurl = {https://ui.adsabs.harvard.edu/abs/2025ApJ...987L...4C}
}

@ARTICLE{2024A&A...691A.172B,
       author = {{Battaglia}, Andrea Francesco and {Krucker}, S{\"a}m and {Veronig}, Astrid M. and {Stiefel}, Muriel Zo{\"e} and {Warmuth}, Alexander and {Benz}, Arnold O. and {Ryan}, Daniel F. and {Collier}, Hannah and {Harra}, Louise},
        title = "{The observational evidence that all microflares that accelerate electrons to high energies are rooted in sunspots}",
      journal = {\aap},
         year = 2024,
        month = nov,
       volume = {691},
          eid = {A172},
        pages = {A172},
          doi = {10.1051/0004-6361/202451152},
archivePrefix = {arXiv},
       eprint = {2409.14466},
 primaryClass = {astro-ph.SR},
       adsurl = {https://ui.adsabs.harvard.edu/abs/2024A&A...691A.172B}
}

@ARTICLE{2013ApJ...765..143I,
       author = {{Ishikawa}, Shin-nosuke and {Krucker}, S{\"a}m and {Ohno}, Masanori and {Lin}, Robert P.},
        title = "{Suzaku/WAM and RHESSI Observations of Non-thermal Electrons in Solar Microflares}",
      journal = {\apj},
         year = 2013,
        month = mar,
       volume = {765},
       number = {2},
          eid = {143},
        pages = {143},
          doi = {10.1088/0004-637X/765/2/143},
       adsurl = {https://ui.adsabs.harvard.edu/abs/2013ApJ...765..143I}
}

@ARTICLE{1961ApJ...133..983N,
       author = {{Newkirk}, Jr., Gordon},
        title = "{The Solar Corona in Active Regions and the Thermal Origin of the Slowly Varying Component of Solar Radio Radiation.}",
      journal = {\apj},
         year = 1961,
        month = may,
       volume = {133},
        pages = {983},
          doi = {10.1086/147104},
       adsurl = {https://ui.adsabs.harvard.edu/abs/1961ApJ...133..983N}
}

@ARTICLE{2020ApJ...905...43C,
       author = {{Chen}, Xingyao and {Kontar}, Eduard P. and {Chrysaphi}, Nicolina and {Jeffrey}, Natasha L.~S. and {Gordovskyy}, Mykola and {Yan}, Yihua and {Tan}, Baolin},
        title = "{Subsecond Time Evolution of Type III Solar Radio Burst Sources at Fundamental and Harmonic Frequencies}",
      journal = {\apj},
         year = 2020,
        month = dec,
       volume = {905},
       number = {1},
          eid = {43},
        pages = {43},
          doi = {10.3847/1538-4357/abc24e},
archivePrefix = {arXiv},
       eprint = {2010.08782},
 primaryClass = {astro-ph.SR},
       adsurl = {https://ui.adsabs.harvard.edu/abs/2020ApJ...905...43C}
}

\end{document}